# Modal interactions between a linear oscillator and a nonlinear absorber: multiscale energy transfers

Lan Huang * Shufeng Lu Wensai Ma

Department of Mechanics, Inner Mongolia University of Technology, Hohhot 010051, PR China

**Abstract:** In aerospace engineering, considerable attention has been given to the use of a nonlinear energy sink (NES) as an efficient passive control device. This paper uses the Complexification-Averaging/Geometric Singular Perturbation Theory (CX-A/GSPT) to investigate the multiscale energy transfers in a forced two-DOF system coupled to a grounded NES (GNES) with both nonlinear grounded stiffness and damping. Due to the coexistence of fundamental resonance and internal resonance in the system, complex modal interactions occur between the linear oscillator (LO) and the GNES. In addition, the ratio of the mass of the GNES and the LO can be seen as a perturbation. This may lead to the occurrence of multiscale dynamics and energy transfers in the system. By using the CX-A method, the slow flow equations of the system can be obtained. Further application of GSPT can obtain the critical manifold that is equivalent to the so-called slow invariant manifold (SIM). With different values of nonlinear grounded stiffness and damping, the critical manifolds have different structures. These critical manifolds can capture diverse types of system responses. Moreover, Hilbert-Huang transform (HHT) is used to analyze the time-frequency-energy relationship of system response. With different parameters, the instantaneous frequencies of the responses of the LO and GNES show a completely different change. The Hilbert spectrums confirm the occurrence of complex modal interactions and different time-scale energy transfers in the system.

E-mail address:
lanhuang0919@163.com( L. Huang, ∗ Corresponding author)
shufenglu@163.com( S.F. Lu)
wensaima@163.com( W.S. Ma)

## 1. Introduction

Vibration is present on various levels in many fields [1-5]. In aerospace engineering, vibration issues often have adverse effects on structural responses and mechanical operations [6-8]. For example, the aerodynamic pulse vibration, resulting from the rotor blades of the first stage of the low-pressure compressor, can induce the vibration fault that may occur at the turbine rotor seals [9]. Under working conditions such as high power density, heavy load and high speed, the traveling wave vibration can cause thin-walled gear failure and even damage that further triggers the breakdown of gearbox and crash of the aircraft [10]. Random vibrations lead to the failure of the airfoil structure of hypersonic airfoil [11]. Hence, the unwanted vibration suppression has been the subject of substantial research interest [12-15].

It has been demonstrated that linear dynamic vibration absorbers can be effective in mitigating unwanted vibrations at a specific frequency [16]. However, most of these linear absorbers are only effective on very narrow frequency bands. Given this, a light nonlinear vibration absorber, namely a nonlinear energy sink (NES), has received much attention. Unlike the linear vibration absorber, due to its essentially nonlinear characteristic, NES is capable of absorbing energy over a relatively wide frequency range from different modes of the main structure [17-20]. This one-way and irreversible (on average) energy flow phenomenon from the linear system to the NES is known as targeted energy transfer (TET) [21]. The TET was found to be a transient resonance capture (TRC) [22]. Furthermore, the dynamical mechanism underlying TET can be realized through three distinct mechanisms, namely fundamental and subharmonic TET and nonlinear beating phenomena [23, 24].

In aerospace engineering, considerable attention has been given to the use of an NES as an efficient passive control device. For example, Lee et al. [25, 26] studied both theoretically and experimentally passive and nonlinear TET in suppressing aeroelastic instability using NES. They showed that NES can partially or even completely suppress aeroelastic instability. Moreover, compared with single-degree-of-freedom NES, they demonstrated that multi-degree-of-freedom NESs can greatly improve the robustness of limit cycle oscillation suppression [27]. Similarly, Tian et al. [28] used NES to mitigate the aeroelastic responses and suppress the wing flutter. Wall et al. [29] studied the effectiveness of NESs on suppressing aeroelastic instability and post-instability response of aircraft wings. Huang et al. [30] investigated the flutter control of a three-dimensional wing using a NES. Notably, NES can be used to control the dynamic instability induced by helicopter ground resonance (HGR). Bergeot et al. [31] analyzed steady-state response regimes of a HGR model coupled to NES and explained four steady-state responses. Furthermore, they [32] investigated the passive control of HGR. It is shown that HGR can be completely suppressed, partially suppressed by periodic response or strongly modulated response, under suitable conditions. In addition, NES is applied to reduce the vibration near critical speeds and the resonance peaks in the misalignment in aero-engine rotor systems [33]. In summary, NES can significantly reduce the vibration of the system of aerospace engineering by transferring the unwanted vibration energy from the vibratory system to NES and dissipating it.

In this paper, a GNES with both nonlinear grounded stiffness and damping is studied. Different types of GNES have been studied and applied. For example, Yao et al. [34] used GNES with piecewise linear stiffness to suppress transient and steady-state vibrations within a rotor system. Geng et al. [35] proposed ground-limited NES (GLNES) and discussed its vibration reduction efficiency via the analytical method, simulation and experiment. Sui et al. [36] investigated the response mechanism and damping effect of the GNES with the inerter. Li et al. [37] studied Hamiltonian dynamics and TET of a two-DOF nonlinear mechanical system coupled to a grounded bistable NES (GBNES). Compared to these studies, this paper investigates the nonlinear modal interactions between the LO and the GNES with both nonlinear grounded stiffness and damping. Moreover, a method, so called the Complexification-Averaging/Geometric Singular Perturbation Theory (CX-A/GSPT), is used to analyze the system coupled to the GNES under resonance condition. Three types of critical manifold are obtained by using CX-A/GSPT. These critical manifolds are equivalent to the so-called slow invariant manifolds (SIM) [38-40]. It can be used to capture the different time scale dynamics of the system. Compared with previous work [31, 41, 42] on NES using GSPT, the impact of changes in the external excitation on system responses is discussed in this paper.

This paper aims to study the multiscale energy transfers of a forced two-DOF system coupled to a GNES. Notably, the mass of GNES is far less than that of LO, which tends to cause the occurrence of different time scale energy transfers in the system, under the interaction between fundamental resonance and internal resonance. In fact, under suitable conditions, the system can perform a strongly modulated response (SMR). It can be seen as periodic or quasi-periodic TET [21]. Especially, when the mass ratio of GNES to LO is a perturbation, SMR can undergo multiscale energy transfer [43]. The numerical results show that the system can take multiscale and intense energy transfers in the system when the critical manifold has a pair of saddle-node bifurcation points. Significantly, Hilbert-Huang transform (HHT) is applied to analyze the time-frequency-energy relationship of several nonstationary responses of the system. Hilbert spectrums show that there are different scale energy transfers in the system when the nonlinear grounded stiffness and damping are fixed at different values.

The outline of this work is as follows. Section 2 describes the mathematical model of the 2-DOF mechanical system and gets the slow flow of the non-dimensional system by using the coordinate transformation and the complexification-averaging technique. Section 3 determines the two fast-slow forms of the slow flow on different time scales. Section 4 discusses in detail the positions of fixed points for the three different types of the critical manifold. The modal interactions and the energy transfers between the LO and the GNES on different time scales are studied in Section 5. Section 6 provides some concluding remarks.

## 2. Model description

The basic model under consideration comprises a two-DOF mechanical system consisting of a linear primary system and a GNES as an absorber. Consider that $x_1$ and $x_2$ are the displacements of the LO and the GNES respectively. The governing equations of the system can be shown as:

$$\begin{cases} m_1\ddot{x}_1 + k_1' x_1 + c_1'\dot{x}_1 + k_2'\left(x_1 - x_2\right) + c_2'\left(\dot{x}_1 - \dot{x}_2\right) = f\cos(\Omega t) \\ m_2\ddot{x}_2 + k_2'\left(x_2 - x_1\right) + c_2'\left(\dot{x}_2 - \dot{x}_1\right) + k' x_2^3 + c' x_2^2 \dot{x}_2 = 0 \end{cases} \tag{1}$$

The LO is supposed to have the mass $m_1$, the damping $c_1'$, and the stiffness $k_1'$. It is subjected to the external force $f\cos(\Omega t)$. The GNES has the very light mass $m_2$, i.e., $m_2/m_1 = \varepsilon$ ($0<\varepsilon<<1$), the nonlinear damping $c'$ and the nonlinear stiffness $k'$. Similarly, $k_2'$ and $c_2'$ are linear coupling stiffness and damping.

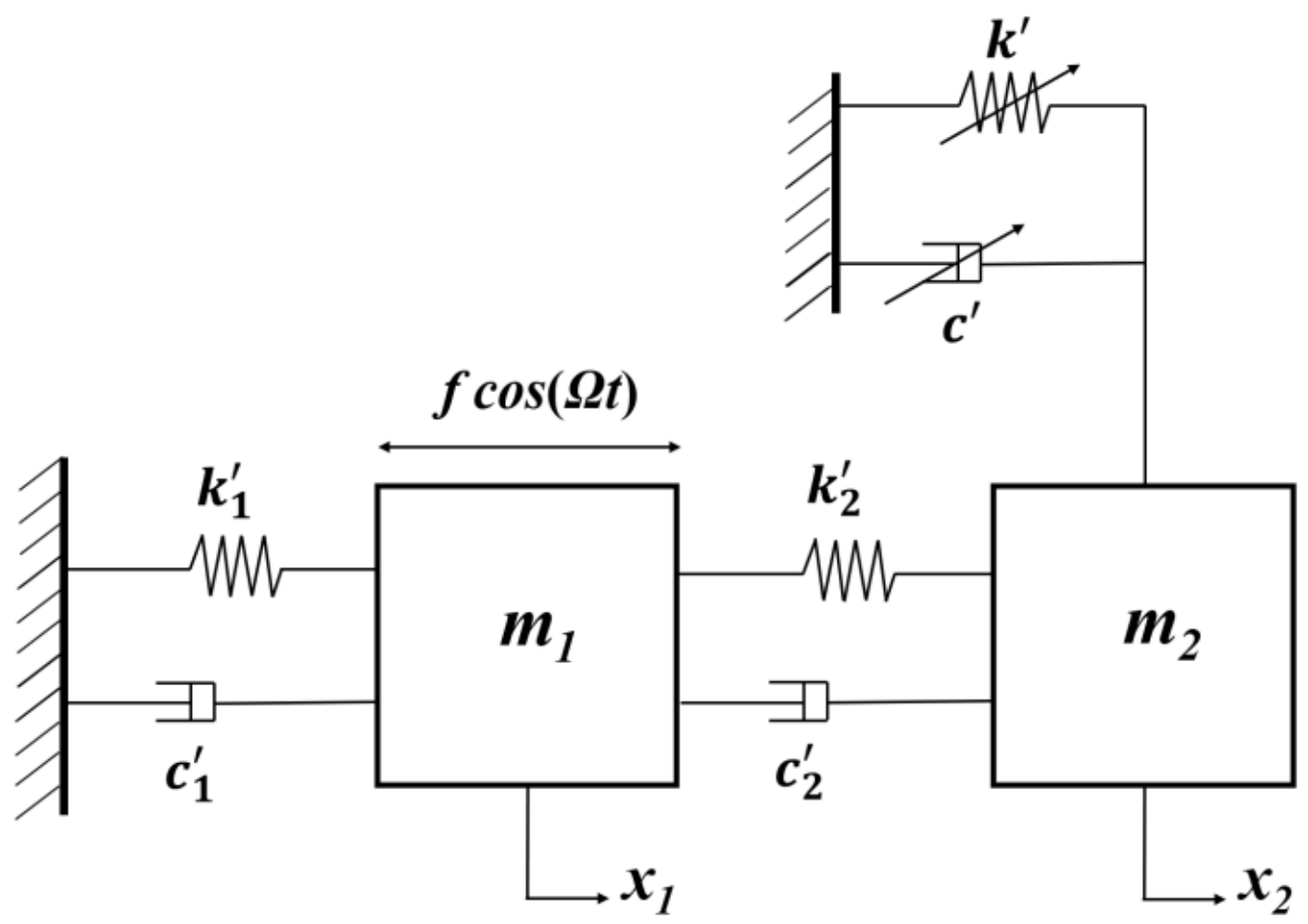

Fig.1. The two-DOF mechanical system consists of a LO under an external excitation coupled to a GNES.

The non-dimensional equations of Eq.(1) read

$$\begin{cases} \ddot{x}_1 + x_1 + \varepsilon\lambda_1\dot{x}_1 + \varepsilon k_2\left(x_1 - x_2\right) + \varepsilon\lambda_2\left(\dot{x}_1 - \dot{x}_2\right) = \varepsilon A\cos\omega\tau \\ \varepsilon\ddot{x}_2 + \varepsilon k_2\left(x_2 - x_1\right) + \varepsilon\lambda_2\left(\dot{x}_2 - \dot{x}_1\right) + \varepsilon k x_2^3 + \varepsilon\lambda x_2^2\dot{x}_2 = 0 \end{cases} \tag{2}$$

where

$$\begin{gathered} \omega_0^2 = \frac{k_1'}{m_1}, \omega_0 t = \tau, \frac{c_1'}{m_1\omega_0} = \varepsilon\lambda_1, \frac{k_2'}{m_1\omega_0^2} = \varepsilon k_2, \frac{c_2'}{m_1\omega_0} = \varepsilon\lambda_2, \\ \frac{k'}{m_1\omega_0^2} = \varepsilon k, \frac{c'}{m_1\omega_0} = \varepsilon\lambda, \frac{f}{m_1\omega_0^2} = \varepsilon A, \omega = \frac{\Omega}{\omega_0}, 0 < \varepsilon \ll 1. \end{gathered} \tag{3}$$

Eq.(4) can be rewritten as

$$\begin{cases} \ddot{x}_1 + x_1 + \varepsilon\lambda_1\dot{x}_1 + \varepsilon k_2\left(x_1 - x_2\right) + \varepsilon\lambda_2\left(\dot{x}_1 - \dot{x}_2\right) = \varepsilon A\cos\omega\tau \\ \ddot{x}_2 + k_2\left(x_2 - x_1\right) + \lambda_2\left(\dot{x}_2 - \dot{x}_1\right) + k x_2^3 + \lambda x_2^2\dot{x}_2 = 0 \end{cases} \tag{4}$$

In the numerical results obtained based on the Runge-Kutta algorithm, Eq.(2) and Eq.(4) are equivalent to each other. Eq.(4) can be seen as a two-DOF system consisting of a LO that is coupled with a strongly nonlinear oscillator. From the perspective of the LO, the linear couplings and nonlinear elements are weak. However, from the perspective of the GNES, the linear couplings and nonlinear elements are strong. This means the occurrence of different time scales coupling, resulting in multiscale dynamics of the system.

Introducing coordinate transformation $x_1 = u_1$, $x_2 = v_1$, Eq.(4) can be written as the four-dimensional real vector fields

$$\begin{cases} \dot{u}_1 = u_2 \\ \dot{u}_2 = -u_1 - \varepsilon\lambda_1 u_2 - \varepsilon k_2 \left(u_1 - v_1\right) - \varepsilon\lambda_2 \left(u_2 - v_2\right) + \varepsilon A \cos\omega\tau \\ \dot{v}_1 = v_2 \\ \dot{v}_2 = -k_2 \left(v_1 - u_1\right) - \lambda_2 \left(v_2 - u_2\right) - k v_1^3 - \lambda v_1^2 v_2 \end{cases} \tag{5}$$

The complex variables [44] are introduced as follows

$$\begin{cases} \psi_1 = u_2 + \mathrm{j}\omega u_1 \\ \psi_1^* = u_2 - \mathrm{j}\omega u_1 \\ \psi_2 = v_2 + \mathrm{j}\omega v_1 \\ \psi_2^* = v_2 - \mathrm{j}\omega v_1 \quad \mathrm{j} = \sqrt{-1} \end{cases} \tag{6}$$

Substituting Eq.(6) into Eq.(5) we get the following form

$$\begin{cases} \dot{\psi}_1 + \dfrac{\varepsilon\lambda_1 - \mathrm{j}\omega}{2}(\psi_1 + \psi_1^*) + \dfrac{1}{2\mathrm{j}\omega}(\psi_1 - \psi_1^*) + \dfrac{\varepsilon k_2}{2\mathrm{j}\omega}(\psi_1 - \psi_1^* - \psi_2 + \psi_2^*) + \dfrac{\varepsilon\lambda_2}{2}(\psi_1 + \psi_1^* - \psi_2 - \psi_2^*) \\ \quad = \dfrac{\varepsilon A}{2}(\exp(\mathrm{j}\omega\tau) + \exp(-\mathrm{j}\omega\tau)) \\ \dot{\psi}_2 - \dfrac{\mathrm{j}\omega}{2}(\psi_2 + \psi_2^*) - \dfrac{k_2}{2\mathrm{j}\omega}(\psi_1 - \psi_1^* - \psi_2 + \psi_2^*) - \dfrac{\lambda_2}{2}(\psi_1 + \psi_1^* - \psi_2 - \psi_2^*) + \dfrac{k}{(2\mathrm{j}\omega)^3}(\psi_2 - \psi_2^*)^3 \\ \quad + \dfrac{\lambda}{2(2\mathrm{j}\omega)^2}(\psi_2 - \psi_2^*)^2(\psi_2 + \psi_2^*) = 0 \end{cases} \tag{7}$$

The periodic steady-state responses of the system (7) can be written in the form where the fast oscillations $\exp(\pm\mathrm{j}\omega\tau)$ is modulated by the slow-varying complex amplitudes $\varphi_i$ ($i$=1,2)

$$\begin{cases} \psi_i = \varphi_i(\tau)\exp(\mathrm{j}\omega\tau), \quad i = 1,2 \\ \psi_i^* = \varphi_i^*(\tau)\exp(-\mathrm{j}\omega\tau), \quad i = 1,2 \end{cases} \tag{8}$$

The derivatives of complex amplitudes $\varphi_i$ ($i$=1,2) with respect to time $\tau$ are obtained

$$\begin{cases} \dot{\psi}_i = \dot{\varphi}_i\exp(\mathrm{j}\omega\tau) + \underbrace{\mathrm{j}\omega\varphi_i\exp(\mathrm{j}\omega\tau)}_{\mathrm{j}\omega\psi_i}, \quad i = 1,2 \\ \dot{\psi}_i^* = \dot{\varphi}_i^*\exp(-\mathrm{j}\omega\tau) - \underbrace{\mathrm{j}\omega\varphi_i^*\exp(-\mathrm{j}\omega\tau)}_{\mathrm{j}\omega\psi_i^*}, \quad i = 1,2 \end{cases} \tag{9}$$

Substituting (8) and (9) into (7), and averaging out the terms that contain fast frequencies higher than $\exp(\mathrm{j}\omega\tau)$,

the slow flow (complex modulation) equations can be obtained

$$\begin{cases} \dot{\varphi}_1 + \left(\dfrac{\varepsilon\lambda_1}{2} + \mathrm{j}\dfrac{\omega^2-1}{2\omega}\right)\varphi_1 + \left(\dfrac{\varepsilon\lambda_2}{2} - \mathrm{j}\dfrac{\varepsilon k_2}{2\omega}\right)(\varphi_1-\varphi_2) = \dfrac{\varepsilon A}{2} \\ \dot{\varphi}_2 + \dfrac{\mathrm{j}\omega}{2}\varphi_2 - \left(\dfrac{\lambda_2}{2} - \mathrm{j}\dfrac{k_2}{2\omega}\right)(\varphi_1-\varphi_2) + \left(\dfrac{\lambda}{8\omega^2} - \mathrm{j}\dfrac{3k}{8\omega^3}\right)|\varphi_2|^2\varphi_2 = 0 \end{cases} \tag{10}$$

## 3. Fast-slow analysis of the slow flow

For the aim of investigating the dynamical behavior around slow flow around the neighborhood of the 1:1:1 resonance regime, where the frequency of external forcing is equal to the frequencies of the fast oscillations $\exp(\pm \mathrm{j}\omega\tau)$, i.e., $\omega=\Omega/\omega_0=1+\varepsilon\sigma$, we, therefore, substitute $\omega=1+\varepsilon\sigma$ into Eq.(10) and obtain

$$\begin{cases} \dot{\varphi}_1 + \left(\dfrac{\varepsilon\lambda_1}{2} + \mathrm{j}\dfrac{(1+\varepsilon\sigma)^2-1}{2(1+\varepsilon\sigma)}\right)\varphi_1 + \left(\dfrac{\varepsilon\lambda_2}{2} - \mathrm{j}\dfrac{\varepsilon k_2}{2(1+\varepsilon\sigma)}\right)(\varphi_1-\varphi_2) = \dfrac{\varepsilon A}{2} \\ \dot{\varphi}_2 + \dfrac{\mathrm{j}(1+\varepsilon\sigma)}{2}\varphi_2 - \left(\dfrac{\lambda_2}{2} - \mathrm{j}\dfrac{k_2}{2(1+\varepsilon\sigma)}\right)(\varphi_1-\varphi_2) + \left(\dfrac{\lambda}{8(1+\varepsilon\sigma)^2} - \mathrm{j}\dfrac{3k}{8(1+\varepsilon\sigma)^3}\right)|\varphi_2|^2\varphi_2 = 0 \end{cases} \tag{11}$$

Based on GSPT [45], Eq.(11) will be written as a fast-slow system in two-dimensional complex vector fields, expressed by

$$\begin{cases} \dfrac{d\varphi_1}{d\tau} = \varepsilon f(\varphi_1, \varphi_2, P, \varepsilon) \\ \dfrac{d\varphi_2}{d\tau} = g(\varphi_1, \varphi_2, P, \varepsilon) \end{cases} \tag{12}$$
$$P = (k, k_2, \lambda, \lambda_1, \lambda_2, \sigma, A)^T$$

where $P$ represents the vector of parameters in Eq.(11). The fast-slow system (12) consists of two coupled subsystems. These two subsystems provide feedback to each other, i.e., modal interactions, resulting in rich dynamical behavior on different time scales. Since $\varepsilon$ is a very small parameter the natural strategy is to decompose solution curves of the fast–slow system into limit segments of the fast subsystem. Given $\varepsilon$=0, we get the fast subsystem

$$\begin{cases} \dfrac{d\varphi_1}{d\tau} = 0 \\ \dfrac{d\varphi_2}{d\tau} = g(\varphi_1, \varphi_2, P, 0) \end{cases} \qquad \text{(Fast subsystem)} \tag{13}$$

In the fast subsystem (13), the complex amplitude $\varphi_1$ can be regarded as a parameter because its derivative with respect to time is zero. When the derivative of complex amplitude $\varphi_2$ in time is equal to zero, i.e., $g(\varphi_1, \varphi_2, P, 0) = 0$, the fast subsystem captures the asymptotic dynamical behavior for the fixed point of the fast-slow system (12). Furthermore, we refer to $\tau$ as the fast time scale and to $\tau^{\wedge}$ as the slow time scale. Setting $\tau = \tau^{\wedge}/\varepsilon$, the equivalent form of the fast-slow system (12) on the slow time scale can be expressed as

$$\begin{cases} \dfrac{d\varphi_1}{d\tau^{\wedge}} = f\left(\varphi_1, \varphi_2, P, \varepsilon\right) \\ \varepsilon \dfrac{d\varphi_2}{d\tau^{\wedge}} = g\left(\varphi_1, \varphi_2, P, \varepsilon\right) \end{cases} \tag{14}$$
$$P = \left(k, k_2, \lambda, \lambda_1, \lambda_2, \sigma, A\right)^T$$

Similarly, one yields the slow subsystem by fixing the small parameter $\varepsilon=0$

$$\begin{cases} \dfrac{d\varphi_1}{d\tau^{\wedge}} = f\left(\varphi_1, \varphi_2, P, 0\right) \\ \quad 0 = g\left(\varphi_1, \varphi_2, P, 0\right) \end{cases} \qquad \text{(Slow subsystem)} \tag{15}$$

It is necessary to note that the slow subsystem (15) is an algebraic-differential equation (ADE). Viewing from the slow subsystem (15), the algebraic equation $g(\varphi_1, \varphi_2, P, 0) = 0$ is a constraint of the evolution of the complex amplitude $\varphi_1$. The fast subsystem (13) and slow subsystem (15) are referred to as the reduced systems because they can be seen as the undisturbed systems corresponding to the disturbed systems (12) and (14) respectively. As $\varepsilon$ is very small ($0<\varepsilon<<1$), the two subsystems are more convenient to study, compared to the fast-slow systems (12) and (14). Especially, the algebraic equation $g(\varphi_1, \varphi_2, P, 0) = 0$, namely the critical manifold, captures the dynamic behavior of the fast-slow systems (12) and (14) approximately. When $g(\varphi_1, \varphi_2, P, 0) = 0$, the critical manifold can be obtained

$$C_0 = \left\{ \left(\varphi_1, \varphi_2\right) \in \mathbb{C}^2 : \left(\lambda_2 - \mathrm{j}k_2\right)\varphi_1 = \left(\lambda_2 - \mathrm{j}\left(k_2 - 1\right)\right)\varphi_2 + \frac{1}{4}\left(\lambda - 3\mathrm{j}k\right)\left|\varphi_2\right|^2 \varphi_2 \right\} \tag{16}$$

## 4. Fixed points and bifurcation structures for the critical manifold

Further, the critical manifold $C_0$, projected on its modulus square plane, can be brought into the following compact form as it follows:

$$C_1 = \left\{ \left(Z_1, Z_2\right) \in \mathbb{R}^{+2} : -\alpha_0 Z_1 = \alpha_1 Z_2 + \alpha_2 Z_2^2 + \alpha_3 Z_2^3 \right\} \tag{17}$$

where $Z_1=|\varphi_1|^2$, $Z_2=|\varphi_2|^2$. Eq. (17) is a cubic polynomial with respect to $Z_2$, where $Z_1$ is a generalized parameter. The coefficients $\alpha_i(i=1,2,3)$ of the cubic polynomial can be written as:

$$\begin{cases} \alpha_0 = \lambda_2^2 + k_2^2 \\ \alpha_1 = -\left(1 + \lambda_2^2 - 2k_2 + k_2^2\right) \\ \alpha_2 = \dfrac{1}{2}\left(-\lambda\lambda_2 + 3k - 3kk_2\right) \\ \alpha_3 = -\dfrac{1}{16}\left(\lambda^2 + 9k^2\right) \end{cases} \tag{18}$$

The critical manifold $C_1$ on the modulus square plane may have degenerated structures due to the influence of generalized parameter $Z_1$ and other system parameters. We especially focus on the effect of generalized parameter $Z_1$ and nonlinear elements $k$ and $\lambda$. As these parameters vary, a folding structure may occur on the critical manifold $C_1$, which causes complex dynamical behavior. By fixing the two group values of system parameters we obtain several folding structures of the critical manifold $C_1$. These structures are plotted in Fig.2.

In Fig.2., we acquire two different structures of the critical manifold $C_1$ with the change of parameters $Z_1$, $k$, and $\lambda$. Eq. (17) has one or three positive roots as the change of these parameters, implying the occurrence of SN bifurcation points. At the bifurcation points, the derivative of the right of the equal sign of Eq. (17) with respect to $Z_2$ should be equal to zero.

$$\alpha_1 + 2\alpha_2 Z_2 + 3\alpha_3 Z_2^2 = 0 \tag{19}$$

Eq.(19) is a bifurcation condition of the occurrence of the SN bifurcation point. Furthermore, when system parameters are fixed at adequate values, $Z_1$ can be seen as the bifurcation parameter in the modulus square plane ($Z_1$, $Z_2$). Therefore, to analyze the number and the positions of SN bifurcation points in the plane ($Z_1$, $Z_2$), we will discuss the following three cases with the different ranges of the coefficients $\alpha_i$($i$=1,2,3) of the cubic polynomial (17).

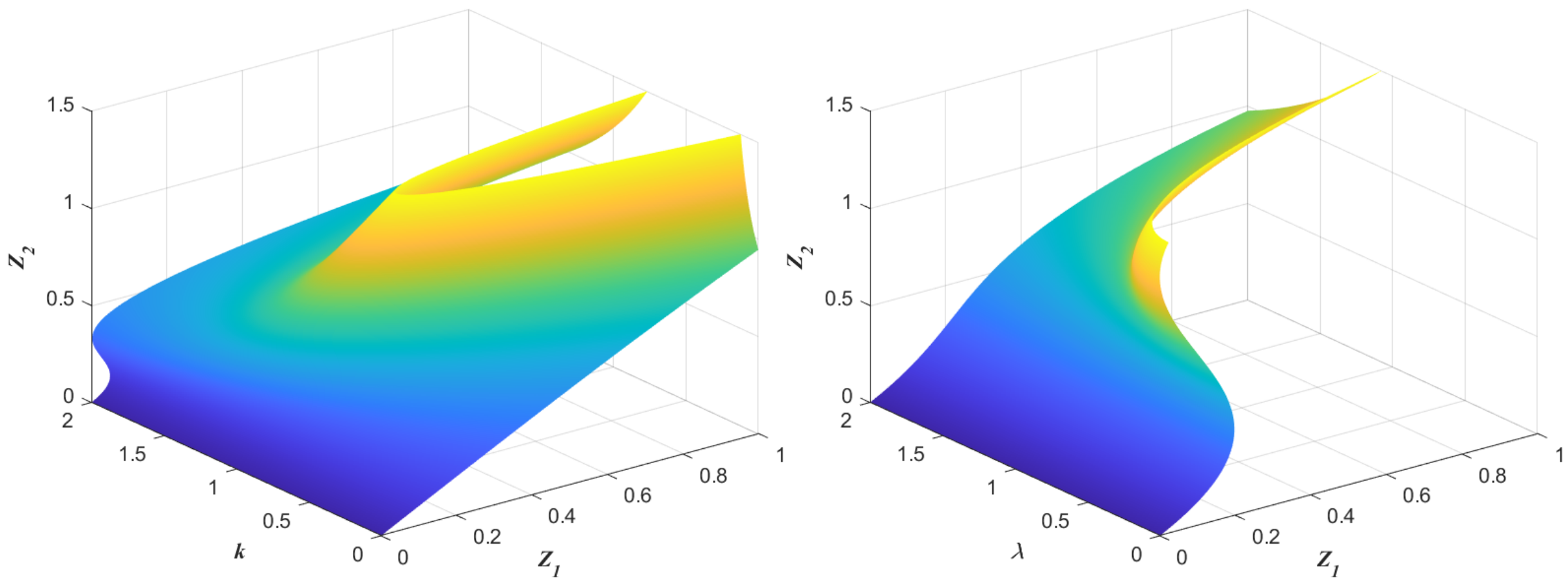

(a) (b)

Fig.2. The critical manifold $C_1$ on the modulus square plane; (a) The critical manifold $C_1$ on the space ($Z_1$, $k$, $Z_2$) with system parameters fixed at $\lambda$=0.5, $\lambda_2$=0, $k_2$=0.5; (b) The critical manifold $C_1$ on the space ($Z_1$, $\lambda$, $Z_2$) with system parameters fixed at $k$ =0.5, $\lambda_2$=0, $k_2$=0.5.

### *4.1.A pair of SN bifurcation points on the critical manifold*

When system parameters satisfy the condition $\alpha_2^2$-3$\alpha_1\alpha_3$>0, there will be a pair of SN bifurcation points with the change of bifurcation parameter $Z_1$, in which the critical manifold $C_1$ contains several folding structures. The two SN bifurcation points are noted as SN$_1$($Z_{11}$, $Z_{21}$) and SN$_2$($Z_{12}$, $Z_{22}$) respectively. According to Eq. (17) and Eq. (19), we obtain the expressions of SN$_1$ and SN$_2$. They are written as Eq. (20) and Eq. (21)

respectively.

$$\mathrm{SN_1}:\begin{cases} Z_{11}=Z_{21}\left\{\dfrac{1}{\lambda_2^2+k_2^2}\left(\left(1+\lambda_2^2-2k_2+k_2^2\right)-\dfrac{1}{2}\left(-\lambda\lambda_2+3k-3kk_2\right)Z_{21}+\dfrac{1}{16}\left(\lambda^2+9k^2\right)Z_{21}^2\right)\right\} \\ Z_{21}=\left\{\begin{array}{l}\dfrac{4}{3\lambda^2+27k^2}\left(\left(-2\lambda\lambda_2+6k\right)-6kk_2\right. \\ \left.+\sqrt{\left(9-27\lambda_2^2-18k_2+9k_2^2\right)k^2-24k\lambda_2\left(1-k_2\right)\lambda+\left(-3+\lambda_2^2+6k_2-3k_2^2\right)\lambda^2}\right)\end{array}\right\} \end{cases} \tag{20}$$

$$\mathrm{SN_2}:\begin{cases} Z_{12}=Z_{22}\left\{\dfrac{1}{\lambda_2^2+k_2^2}\left(\left(1+\lambda_2^2-2k_2+k_2^2\right)-\dfrac{1}{2}\left(-\lambda\lambda_2+3k-3kk_2\right)Z_{22}+\dfrac{1}{16}\left(\lambda^2+9k^2\right)Z_{22}^2\right)\right\} \\ Z_{22}=\left\{\begin{array}{l}-\dfrac{4}{3\lambda^2+27k^2}\left(\left(2\lambda\lambda_2-6k\right)+6kk_2\right. \\ \left.+\sqrt{\left(9-27\lambda_2^2-18k_2+9k_2^2\right)k^2-24k\lambda_2\left(1-k_2\right)\lambda+\left(-3+\lambda_2^2+6k_2-3k_2^2\right)\lambda^2}\right)\end{array}\right\} \end{cases} \tag{21}$$

Fixing system parameters at $k$=0.5, $\lambda$=0.5, $\lambda_2$=0, and $k_2$=0.5 one can get the critical manifold on the modulus square plane ($Z_1$, $Z_2$), in which $Z_1$ is the bifurcation parameter with respect to the variable $Z_2$. As $Z_1$ varies slowly, the number of fixed points as well as the stability of $Z_2$ will change. In Fig.3, the equations $Z_1$=$Z_{11}$ and $Z_1$=$Z_{12}$ divide the modulus square plane ($Z_1$, $Z_2$) into three regions, namely two single-fixed point regions and one three-fixed point region. When bifurcation parameter $Z_1$>$Z_{11}$ or 0<$Z_1$<$Z_{12}$, there is only one stable fixed point on the critical manifold $C_1$. However, as the value of the parameter $Z_1$ is in the interval $Z_{12}$<$Z_1$<$Z_{11}$, the critical manifold $C_1$ will have a pair of stable fixed points and an unstable fixed point. Especially, the two stable critical manifold branches will merge with the one unstable critical manifold at the two bifurcation points $SN_1$ and $SN_2$ respectively. The above analysis is clear from the results of Fig.3.

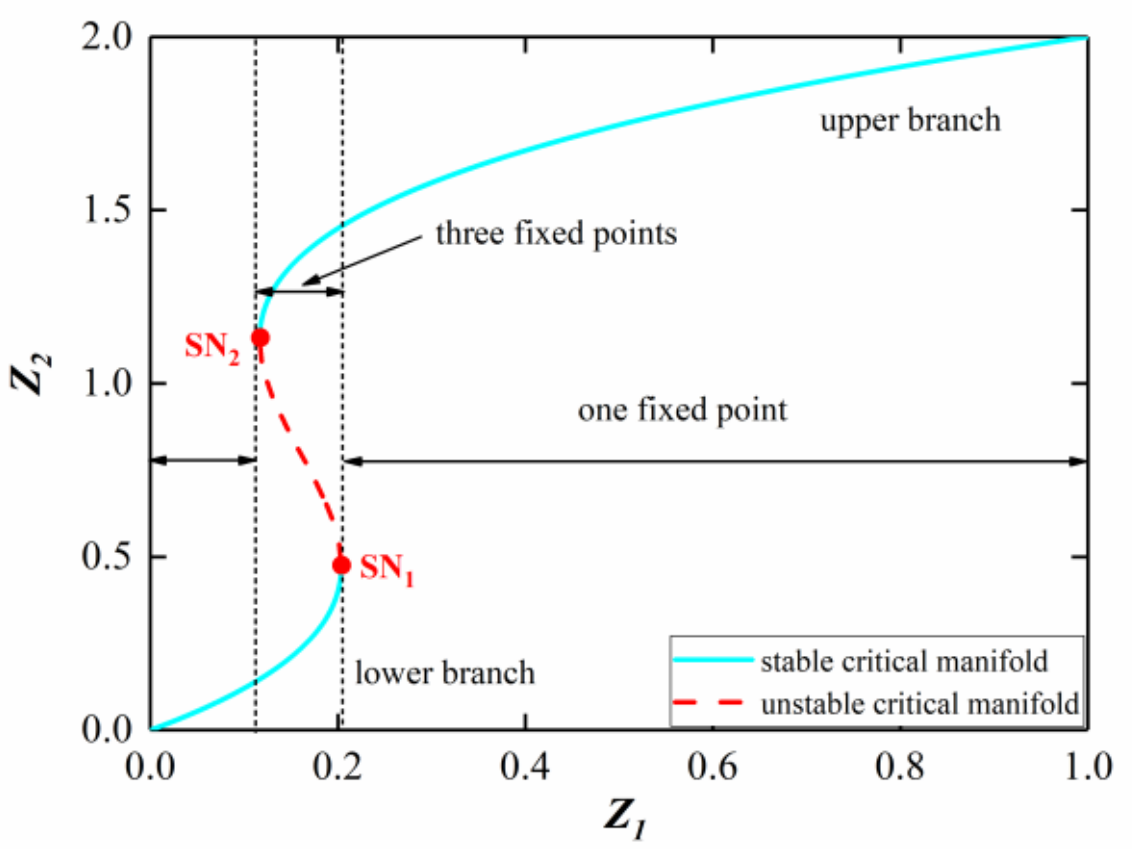

Fig.3. The critical manifold $C_1$ on the modulus square plane ($Z_1$, $Z_2$) with system parameters fixed at $k$=0.5, $\lambda$=0.5, $\lambda_2$=0, $k_2$=0.5.

## *4.2. Tangent point on the critical manifold*

When the values of system parameters satisfy the condition $\alpha_2^2$-3$\alpha_1\alpha_3$=0, the critical manifold on the modulus square plane ($Z_1$, $Z_2$) contains one tangent point, namely TP, that is the tangent of the upper branch and the lower branch of the stable critical manifold. This point is distinct from the SN bifurcation point because there are no changes in the number and stability of the fixed point. However, the regimes around the TP point may cause nontrivial dynamical behavior. Similarly, by combining Eq. (17) and Eq. (19), we get the expression for TP

$$\text{TP:}\quad \begin{cases} Z_1 = Z_2 \left\{ \dfrac{1}{\lambda_2^2+k_2^2}\left( \left(1+\lambda_2^2-2k_2+k_2^2\right) - \dfrac{1}{2}\left(-\lambda\lambda_2+3k-3kk_2\right)Z_2 + \dfrac{1}{16}\left(\lambda^2+9k^2\right)Z_2^2 \right) \right\} \\ Z_2 = \left\{ \dfrac{\left(-24k_2+24\right)k - 8\lambda_2\lambda}{27k^2+3\lambda^2} \right\} \end{cases} \tag{22}$$

In Fig.4, we find that there is only one fixed point on the whole critical manifold with a change of control parameter $Z_1$. Quite remarkably, the unstable critical manifold disappears, while TP appears, which leads to new dynamical behavior.

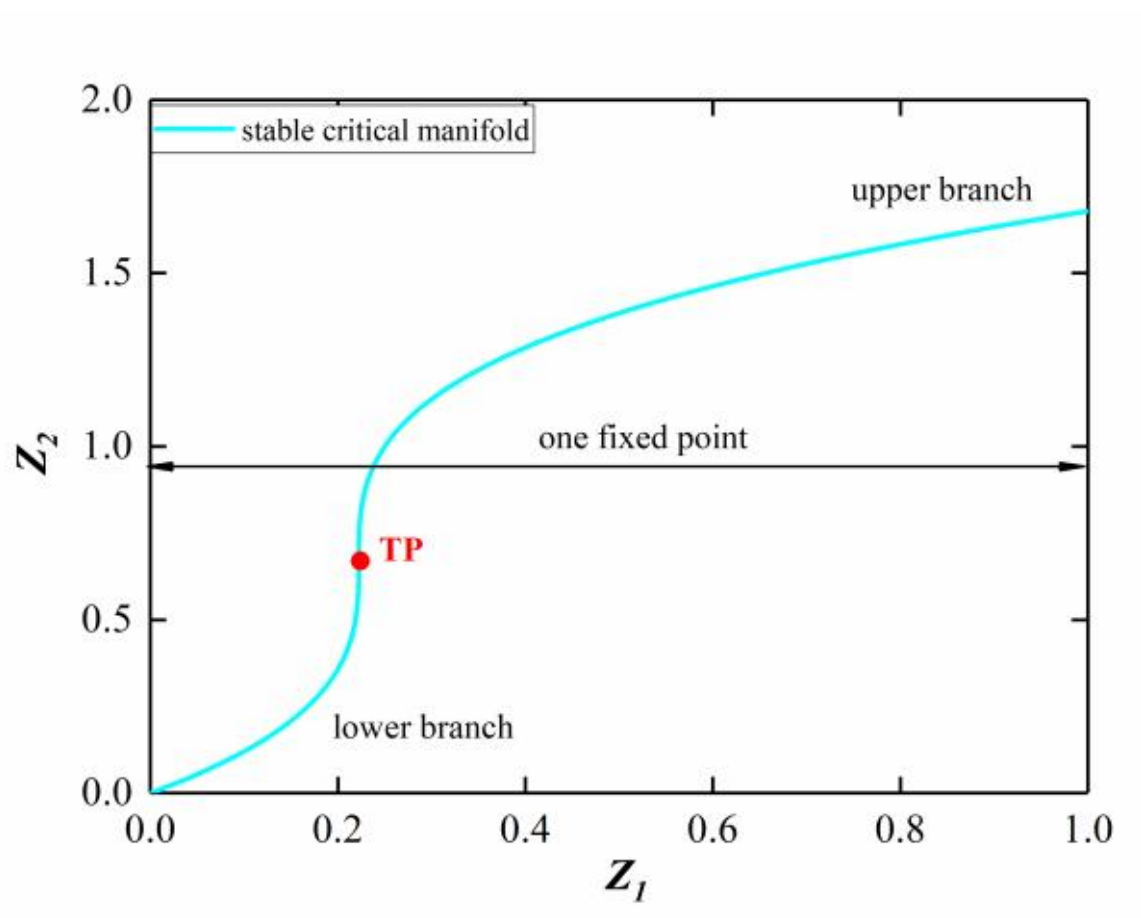

Fig.4. The critical manifold $C_1$ on the modulus square plane ($Z_1$, $Z_2$) with system parameters fixed at $k$=0.5, $\lambda$=0.866, $\lambda_2$=0, $k_2$=0.5.

*4.3. Trivial fixed points on the critical manifold*

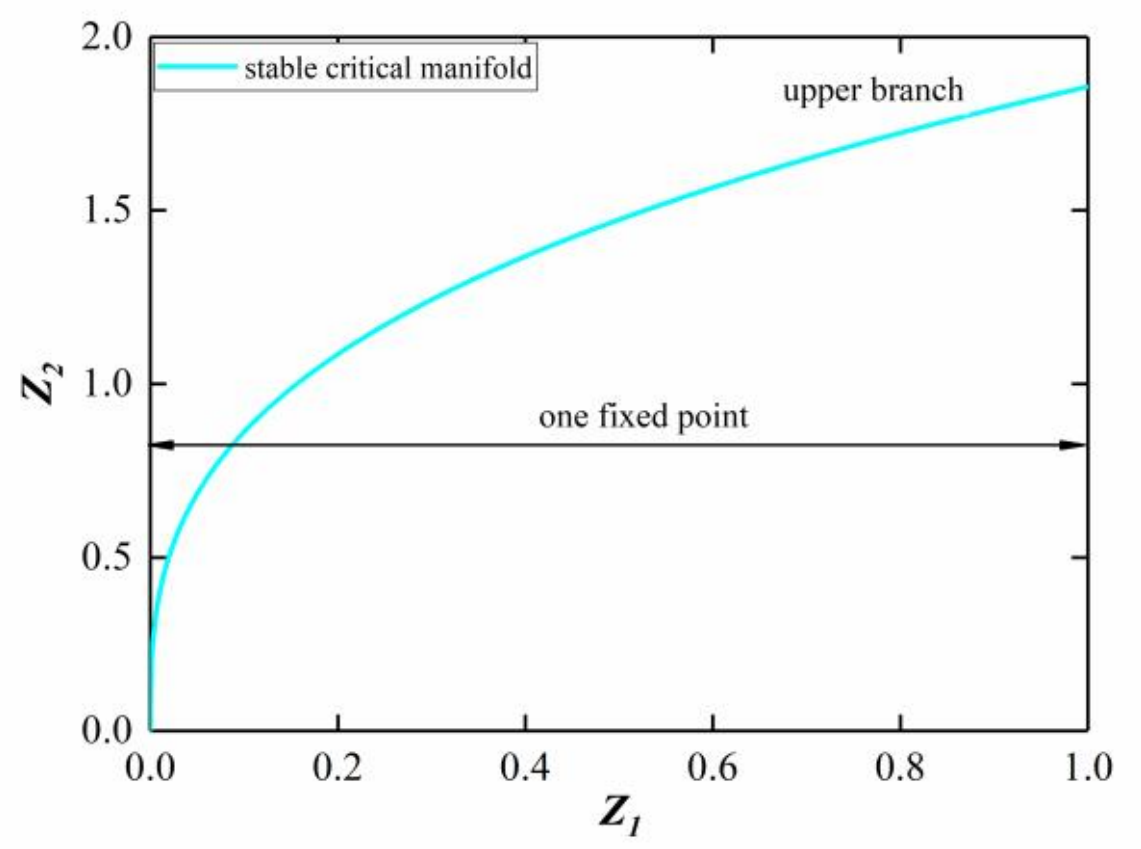

Fig.5. The critical manifold $C_1$ on the modulus square plane ($Z_1$, $Z_2$) with system parameters fixed at $k$=0.5, $\lambda$=0.5, $\lambda_2$=0, $k_2$=1.

When the values of system parameters meet the condition $\alpha_2^2$-3$\alpha_1\alpha_3$<0, there is no special point, i.e., SN

bifurcation point and tangent point, on the critical manifold. Usually, it means that the regimes around the critical manifold do have not any nontrivial dynamical behavior. The case is clear in Fig.5.

### *4.4. Positions of fixed points on the critical manifold*

Introducing the polar coordinates

$$\begin{cases} \varphi_1 = N_1 \mathrm{e}^{\mathrm{j}\theta_1} \\ \varphi_2 = N_2 \mathrm{e}^{\mathrm{j}\theta_2} \qquad N_i \in \mathbb{R}^+, \theta_i \in \mathbb{R}\left(i=1,2\right) \end{cases} \tag{23}$$

where $N_i$ and $\theta_i$ ($i$=1,2) are module and phase angles in polar coordinates. Substituting Eq.(23) into Eq.(16) and separating real and imaginary parts we get

$$\begin{cases} F_1 = \frac{1}{4}\left(-3N_2{}^3 k + \left(-4k_2+4\right)N_2\right)\sin\theta_2 + \frac{1}{4}\left(-\lambda N_2{}^3 - 4N_2\lambda_2\right)\cos\theta_2 + N_1\left(\sin\theta_1 k_2 + \lambda_2\cos\theta_1\right) \\ F_2 = \frac{1}{4}\left(3N_2{}^3 k + \left(4k_2-4\right)N_2\right)\cos\theta_2 + \frac{1}{4}\left(-\lambda N_2{}^3 - 4N_2\lambda_2\right)\sin\theta_2 + N_1\left(\sin\theta_1\lambda_2 - k_2\cos\theta_1\right) \end{cases} \tag{24}$$

Let $F_1$=0 and $F_2$=0 satisfy at the same time, and we obtain the positions of the fixed points, where two curves corresponding to the equations $F_i$=0 ($i$=1,2) intersect each other, on the critical manifold by fixing the values of system parameters. As system parameters are taken at distinct values, the critical manifold may generate different structures, which leads to the change of the positions of fixed points. According to the different values of parameters, thus, we discuss the following three cases.

#### *4.4.1. Case 1*

For $\alpha_2^2$-3$\alpha_1\alpha_3$>0, there are a pair of SN bifurcation points on the critical manifold. In subfigure (a) of Fig.6 one can find that the lower branch of the stable critical manifold and the unstable critical manifold coalesce into the $\mathrm{SN}_1$ bifurcation point. Similarly, the upper branch of the stable critical manifold merges with the unstable critical manifold at the $\mathrm{SN}_2$ bifurcation point. Taking $N_1$=0.1, $N_1$=0.4, and $N_1$=0.7, and fixing the phase angle of the complex amplitude of the LO at $\theta_1$=1, we get five fixed points, namely $\mathrm{P}_i$($i$=1,2,3,4,5), on the plane ($\theta_2$, $N_2$). Among them, $\mathrm{P}_1$, $\mathrm{P}_2$, $\mathrm{P}_4$, and $\mathrm{P}_5$ are stable points on the stable critical manifold, while $\mathrm{P}_3$ is an unstable point on the unstable critical manifold. These positions of the fixed points are shown in subfigure (b)-(d) of Fig.6.

#### *4.4.2. Case 2*

For $\alpha_2^2$-3$\alpha_1\alpha_3$=0, there only is one tangent point, i.e., TP, on the critical manifold. Similarly, we take $N_1$=0.1, $N_1$=0.471, and $N_1$=0.7, and acquire the positions of three fixed points on the plane ($\theta_2$, $N_2$) with the other system parameters fixed at $k$=0.5, $\lambda$=0.866, $\lambda_2$=0, $k_2$=0.5. It is worth noting that the two curves $F_1$=0 and $F_2$=0 are tangent at point TP, which leads to the disappearance of the folding structure and results in the whole critical manifold having only a stable single fixed point at each point. Detailed information with respect to the positions of three fixed points can be seen in subfigure (b)-(d) of Fig.7.

### *4.4.3. Case 3*

For $\alpha_2^2$-3$\alpha_1\alpha_3$<0, there is no SN bifurcation point and TP point on the critical manifold. Exactly, both the lower branch of the stable critical manifold and the unstable critical manifold disappear. At this moment there is only one stable fixed point at each position of the critical manifold. For instance, taking $N_1$=0.5 we obtain the position of the fixed point of the upper branch of the critical manifold on the plane ($\theta_2$, $N_2$), which is depicted in subfigure (b) of Fig.8.

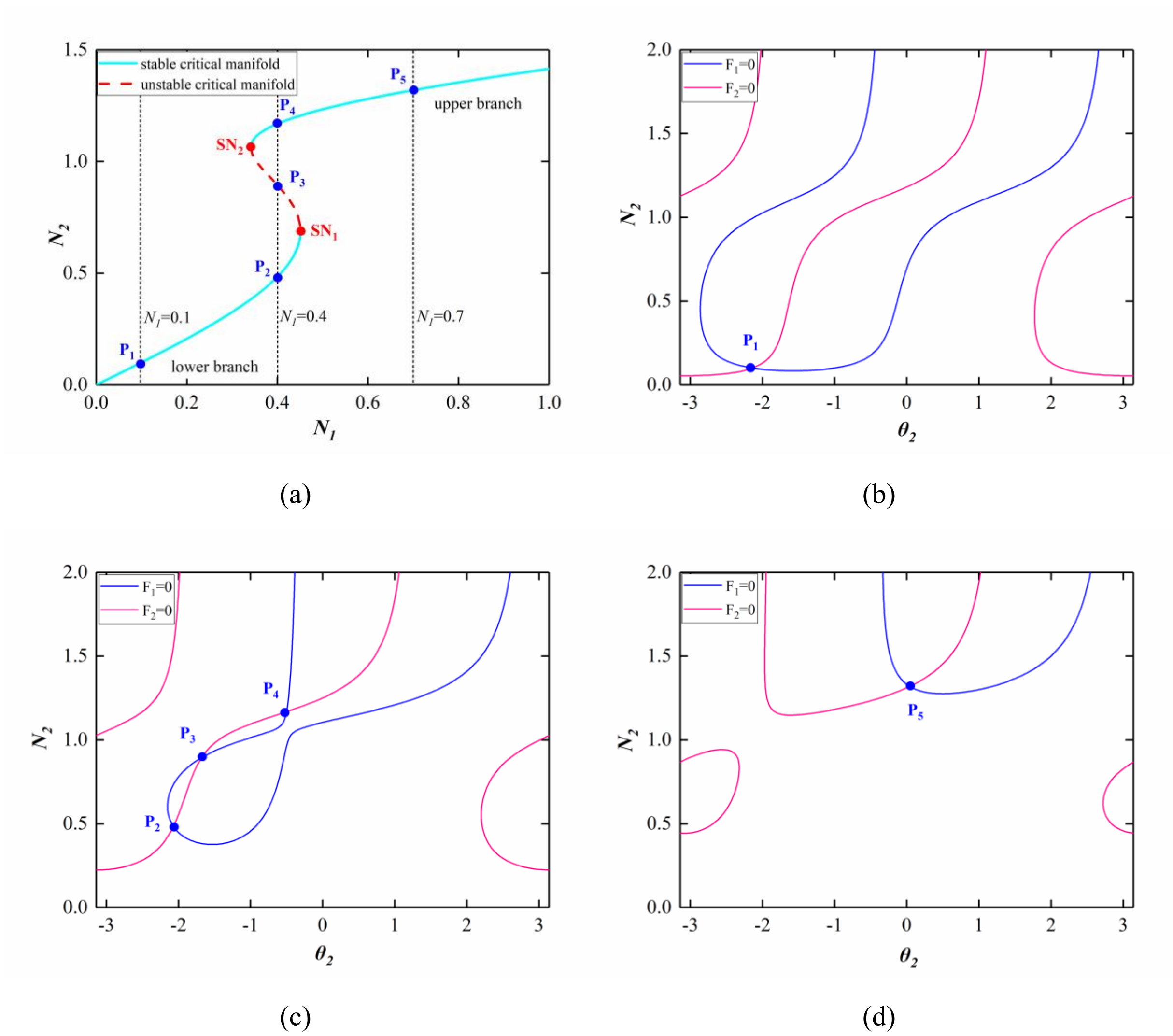

Fig.6. The critical manifold $C_0$ on the modulus plane ($N_1$, $N_2$) and positions of fixed points on the plane ($\theta_2$, $N_2$) with system parameters fixed at $k$=0.5, $\lambda$=0.5, $\lambda_2$=0, $k_2$=0.5;(a) The critical manifold;(b) $N_1$=0.1; (c) $N_1$=0.4; (d) $N_1$=0.7.

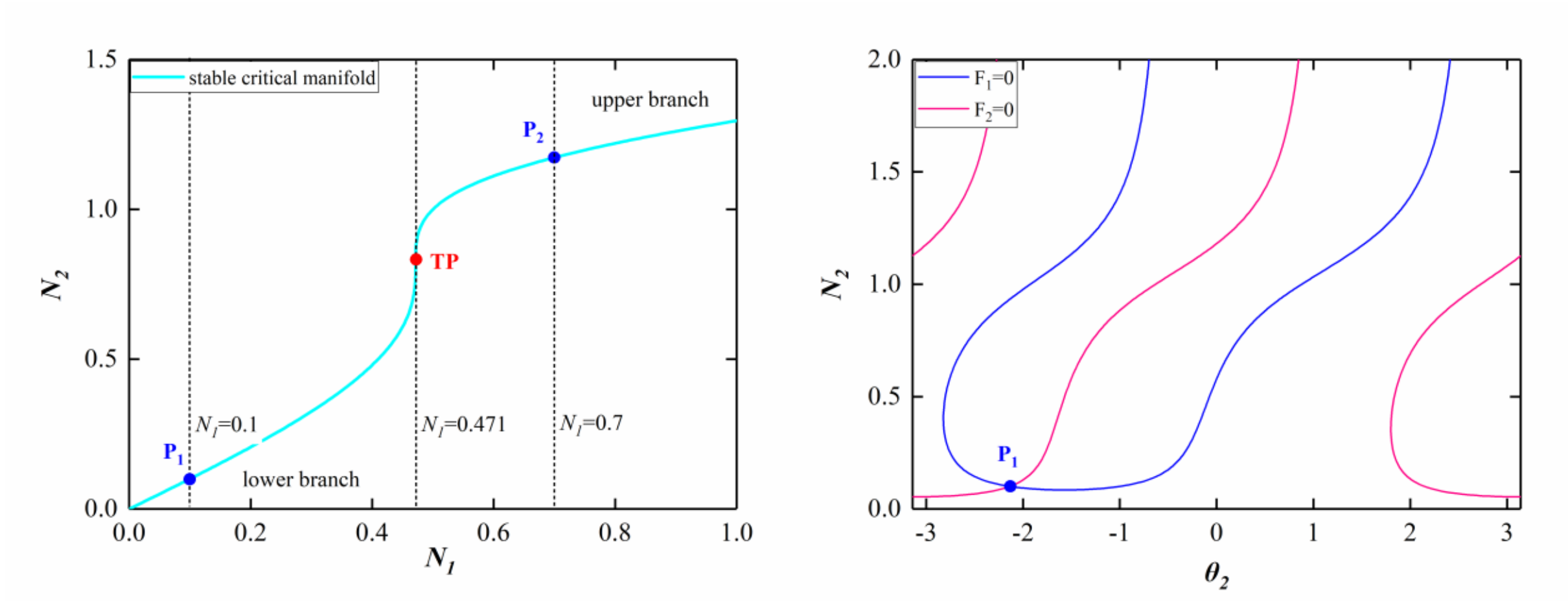

(a) (b)

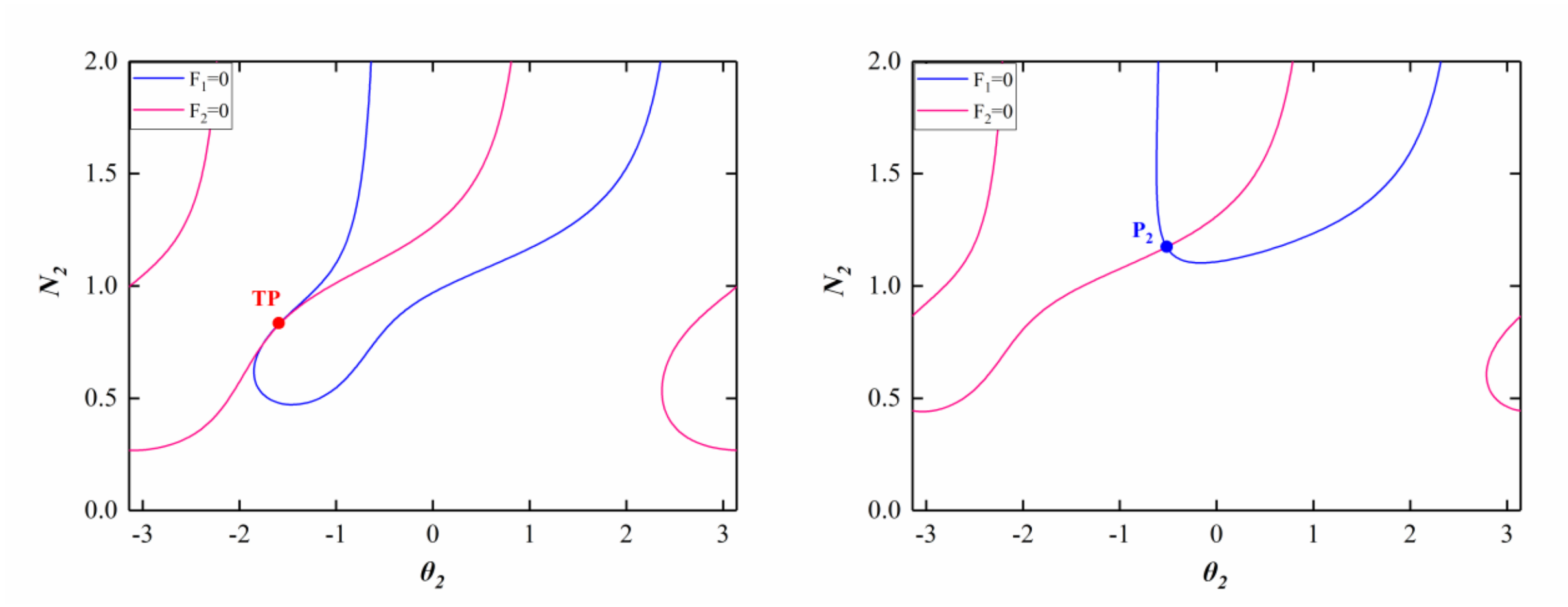

(c) (d)

Fig.7. The critical manifold $C_0$ on the modulus plane ($N_1$, $N_2$) and positions of fixed points on the plane ($\theta_2$, $N_2$) with system parameters fixed at $k$=0.5, $\lambda$=0.866, $\lambda_2$=0, $k_2$=0.5;(a) The critical manifold;(b) $N_1$=0.1; (c) $N_1$=0.471; (d) $N_1$=0.7.

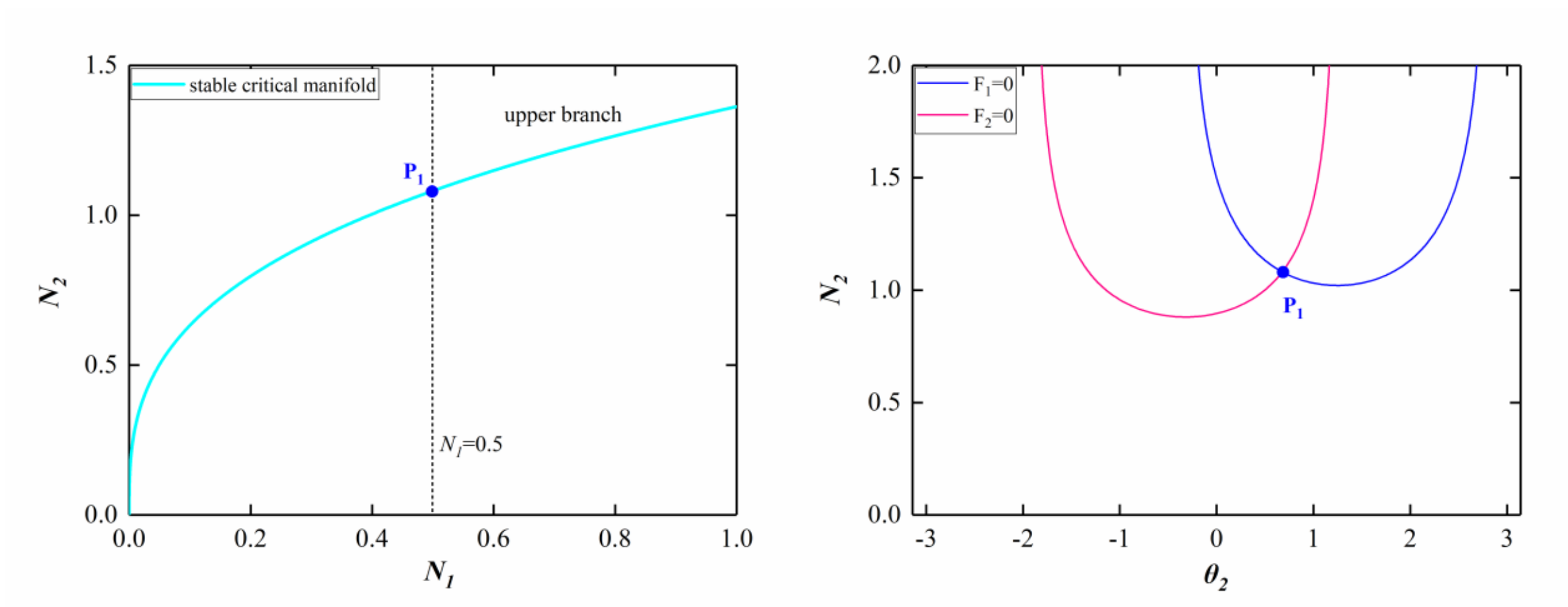

(a) (b)

Fig.8. The critical manifold $C_0$ on the modulus plane ($N_1$, $N_2$) and positions of fixed points on the plane ($\theta_2$, $N_2$) with system parameters fixed at $k$=0.5, $\lambda$=0.5, $\lambda_2$=0, $k_2$=1;(a) The critical manifold;(b) $N_1$=0.5.

### *4.5. Phase relation between $\varphi_1$ and $\varphi_2$ on the critical manifold*

The critical manifold (16) can further describe the phase relationship between the complex amplitudes $\varphi_1$ and $\varphi_2$. Substituting the polar coordinates (23) into the critical manifold (16), we get the following equation

$$N_1 e^{\theta_1} = Y_1 Y_2 N_2 e^{\theta_2} \tag{25}$$

with

$$\begin{cases} Y_1 = r_1 e^{\vartheta_1} \\ r_1 = \left(\lambda_2^2 + k_2^2\right)^{-\frac{1}{2}} \\ \vartheta_1 = \arctan\left(\dfrac{k_2}{\lambda_2}\right) \end{cases} \tag{26}$$

and

$$\begin{cases} Y_2 = r_2 e^{\vartheta_2} \\ r_2 = \left(\left(\lambda_2 + \dfrac{\lambda N_2^2}{4}\right)^2 + \left(1 - k_2 - \dfrac{3kN_2^2}{4}\right)^2\right)^{\frac{1}{2}} \\ \vartheta_2 = \arctan\left(\left(1 - k_2 - \dfrac{3kN_2^2}{4}\right)\left(\lambda_2 + \dfrac{\lambda N_2^2}{4}\right)^{-1}\right) \end{cases} \tag{27}$$

Then, we obtain the phase difference between $\theta_1$ and $\theta_2$

$$\Delta = \theta_1 - \theta_2 = \vartheta_1 + \vartheta_2 \tag{28}$$

Since the fixed point, where $N_2$ is a constant, corresponds to the stationary oscillation of the system, both phases $\vartheta_1$ and $\vartheta_2$ are constant values. Then, we get

$$\frac{d\theta_1}{d\tau_1} = \frac{d\theta_2}{d\tau_1} = \varpi, \quad \varpi \in \mathbb{R} \tag{29}$$

Consequently, the periods of evolution along the critical manifold for the two complex amplitudes $\varphi_1$ and $\varphi_2$ are equal. The period is given as:

$$\mathrm{T}_{\varphi} = \mathrm{T}_{\varphi_1} = \mathrm{T}_{\varphi_2} = \frac{\pi}{\varpi} \tag{30}$$

## 5. Modal interactions between the two oscillators on different time scales

Around the neighborhood of the 1:1:1 resonance regime, the two-DOF system (2) will occur modal interactions between the LO and the GNES, which results in the different time scales coupled oscillations. We can find from the slow flow (11) that the complex amplitude $\varphi_1$ of the LO is a very slow variable relative to the complex amplitude $\varphi_2$ of the GNES. This means, at the same time, the number of GNES vibrations is much greater than that of the LO vibrations. More precisely, the complex amplitude $\varphi_1$ of the LO is a slow-varying variable on the critical manifold (16). Therefore, $N_1$ can be seen as a generalized parameter of the critical manifold on the modulus plane ($N_1$, $N_2$). In addition, with the change of $N_1$, the critical manifold may occur SN bifurcation points causing the folding structures and TP point connecting the upper and the lower stable critical manifold. All of these structures of the critical manifold obtained by qualitative analysis imply that the distinct types of complex modal interactions and energy transfers between the LO and the GNES may occur. Consequently, by using the Hilbert-Huang transform, we get Hilbert spectrums corresponding to oscillations on different time scales and obtain the instantaneous frequencies occurring in the diverse types of modal interactions.

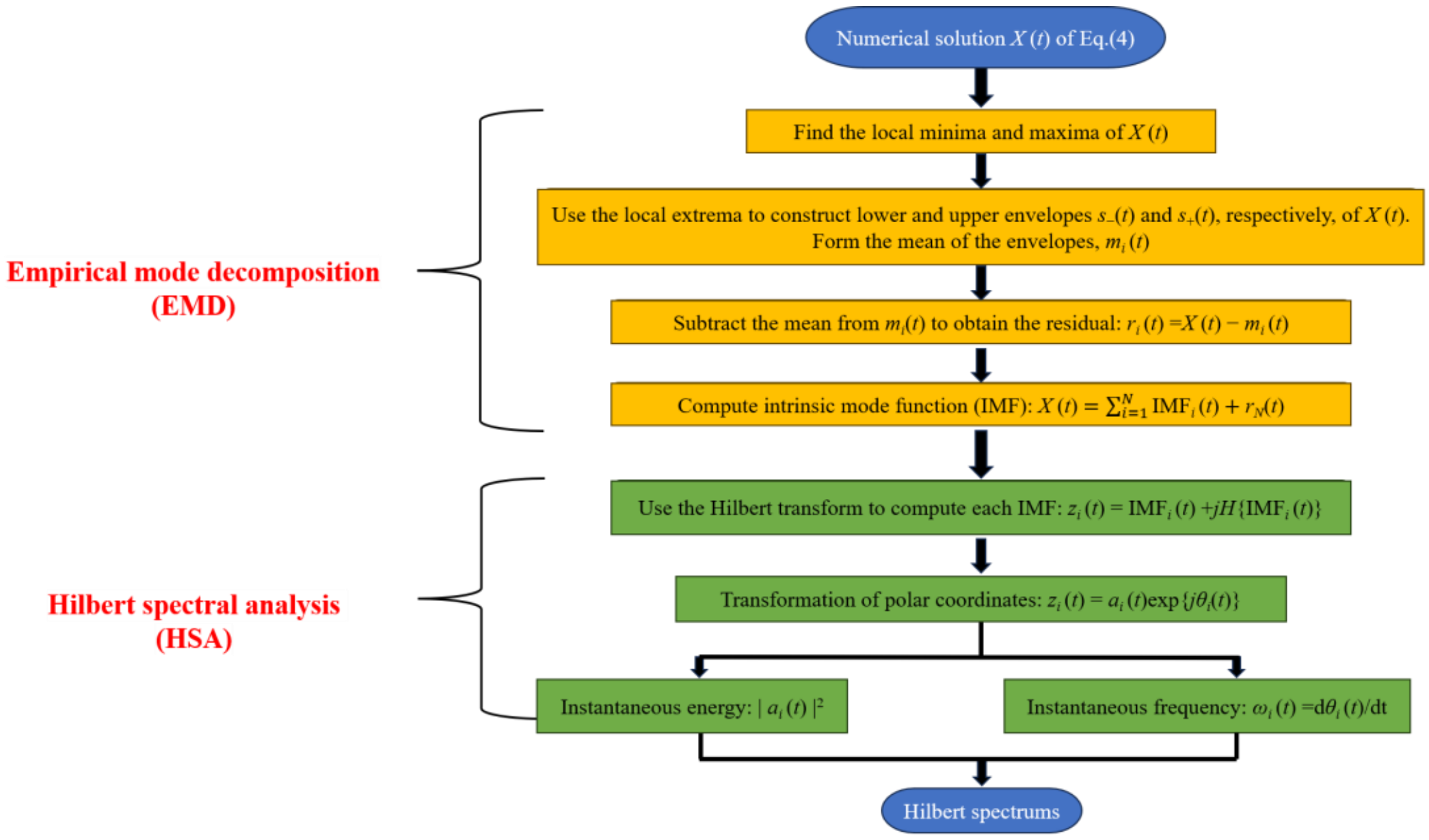

Fig.9. The flow chart for the Hilbert spectrums

HHT consists of two parts: empirical mode decomposition (EMD) and Hilbert spectral analysis (HSA). This method is potentially viable for nonlinear and nonstationary data analysis, especially for time-frequency-energy representations [46]. It has been tested and validated exhaustively, but only empirically. In all the cases studied, the HHT gave results much sharper than those from traditional analysis methods in time-frequency-energy representations. The Hilbert spectrums show the instantaneous frequency and energy with the change of time. Accordingly, the flow chart for the Hilbert spectrums is shown in Fig.9.

For the sake of brevity, we fix the initial point at the origin (0,0,0,0) and take the two linearly couple damping equal to zero, i.e., $\lambda_1$=0, $\lambda_2$=0 during numerical simulations obtained from direct integration of Eq.(2). The Runge–Kutta scheme has been used thanks to the ode45 function of Matlab.

*5.1. Evolution of oscillations on three-time-scale*

Taking system parameters at $\varepsilon$=0.001, $k$=1, $\lambda$=0.5, $k_2$=0.7, $\sigma$=4, $A$=0.5, we get numerical results obtained from direct numerical integration of Eq.(2) on the modulus plane ($N_1$, $N_2$), superposed by the critical manifold (16). Viewing from subfigure (a) of Fig.10, we can find that the critical manifold approximately predicts the trajectory of the system (2). Indeed, the trajectory of the system starting from the origin moves along the lower branch of the critical manifold until it reaches the $SN_1$ bifurcation point, and then it jumps to the upper branch of the critical manifold. Further, driven by the external excitation, the trajectory oscillates along the upper branch of the critical manifold until it reaches the $SN_2$ point, and it jumps to the lower branch of the critical manifold. So far, the trajectory of the system completes a period of motion. After that, the evolution of the system follows an approximate trajectory.

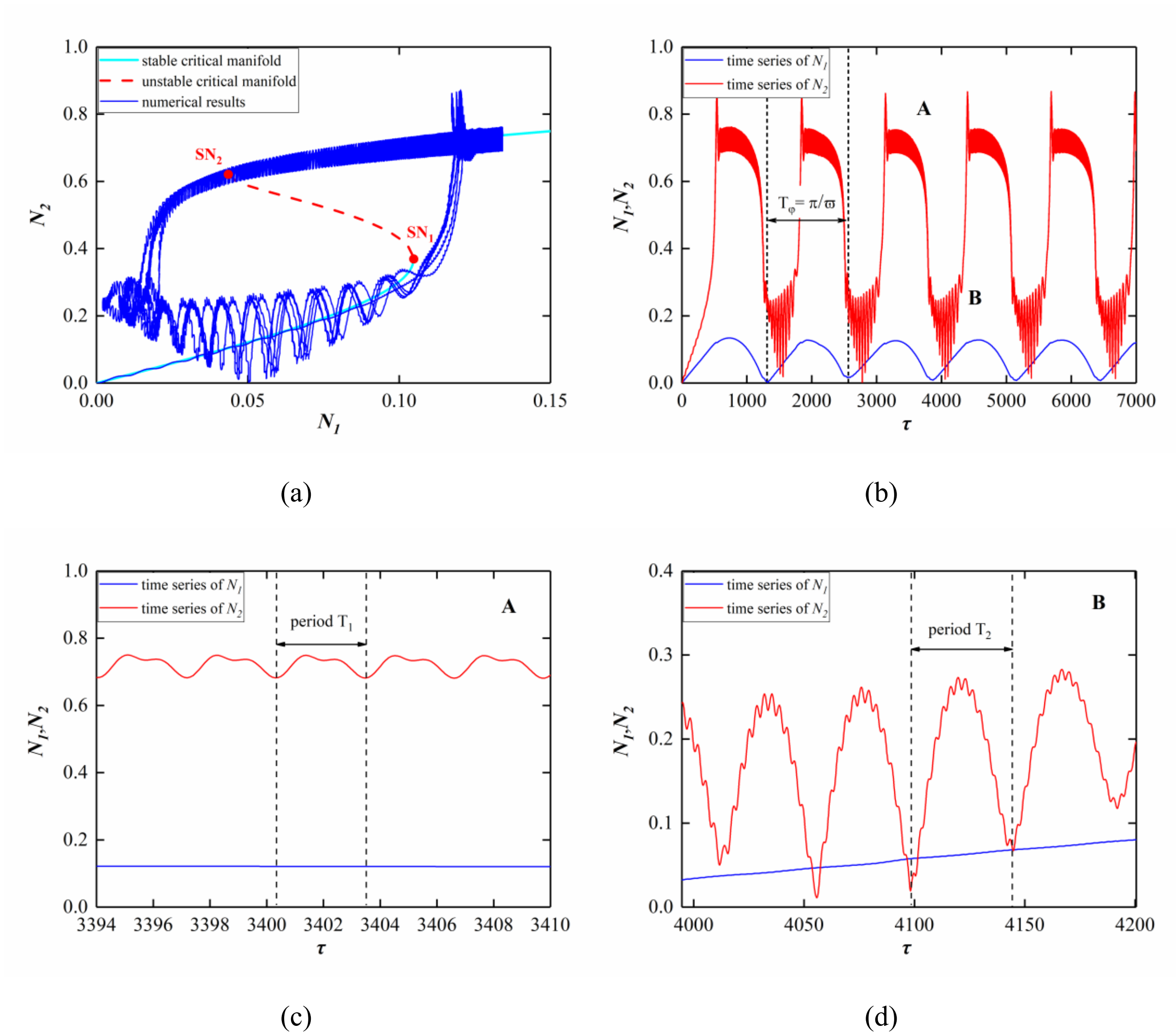

Fig.10. Numerical simulations with the system parameters fixed at $\varepsilon$=0.001, $k$=1, $\lambda$=0.5, $k_2$=0.7, $\sigma$=4, $A$=0.5;(a) Superposition of the critical manifold on the modulus plane ($N_1$, $N_2$) and corresponding numerical results;(b) Superposition of numerical results of the evolution of $N_1$ and $N_2$;(c) Close-up A; (d) Close-up B.

From the subfigure (b) of Fig.10, one can acquire the time series simulation of modulus (i.e., $N_1$ and $N_2$)

of complex amplitudes of the LO and the GNES. We notice that the time series of the evolution of $N_1$ and $N_2$ approximately exhibit regular periodic motion along the critical manifold, and the period of both movements is equal to $T_\varphi=\pi/\varpi$. However, the time series of the evolution of $N_2$ presents also nontrivial periodic motion, including two jumping processes and fast oscillations. In subfigure (c) of Fig.10 one can find the oscillation of $N_2$ on the upper branch of the critical manifold is much faster than that of $N_1$. One can realize that $N_1$ remains unchanged during one oscillation period (i.e., $T_1$) of fast oscillations for $N_2$. However, on the lower branch of the critical manifold, there is a slight change in $N_1$ during one oscillation period $T_2$ of fast oscillations for $N_2$. Similarly, we can realize a similar situation from subfigure (d) of Fig.10.

Consequently, under the excitation of external forcing, the modal of the two oscillators periodically reacts with each other, which leads to nontrivial oscillations along the critical manifold. Due to $T_1 \ll T_2 \ll T_\varphi$, therefore, we define the oscillations of this pattern as three-time-scale oscillations. Applying the Hilbert-Huang transform to $x_1$ and $x_2$ corresponding to the displacements for the LO and the GNES, we get the instantaneous frequencies with the change of time $\tau/2\pi$. In the subfigure (a) of Fig.11, the Hilbert spectrum, consisting of instantaneous frequency $\omega_{\text{LO}}$, time $\tau/2\pi$, and instantaneous energy that is represented by the color bar, indicates that $\omega_{\text{LO}}$ is nearly equal to 1 throughout the entire time range and remains unchanged. However, from Fig.11(b), we observe two regions, denoted as interval (0.95,1.05) and interval (0.85,1), of significant changes in the instantaneous frequency $\omega_{\text{NES}}$. Further, one finds that $\omega_{\text{NES}}$ oscillates rapidly in intervals (0.95,1.05) and slowly changes in intervals (0.85,1), which implies the occurrence of the modal interactions on different time scales.

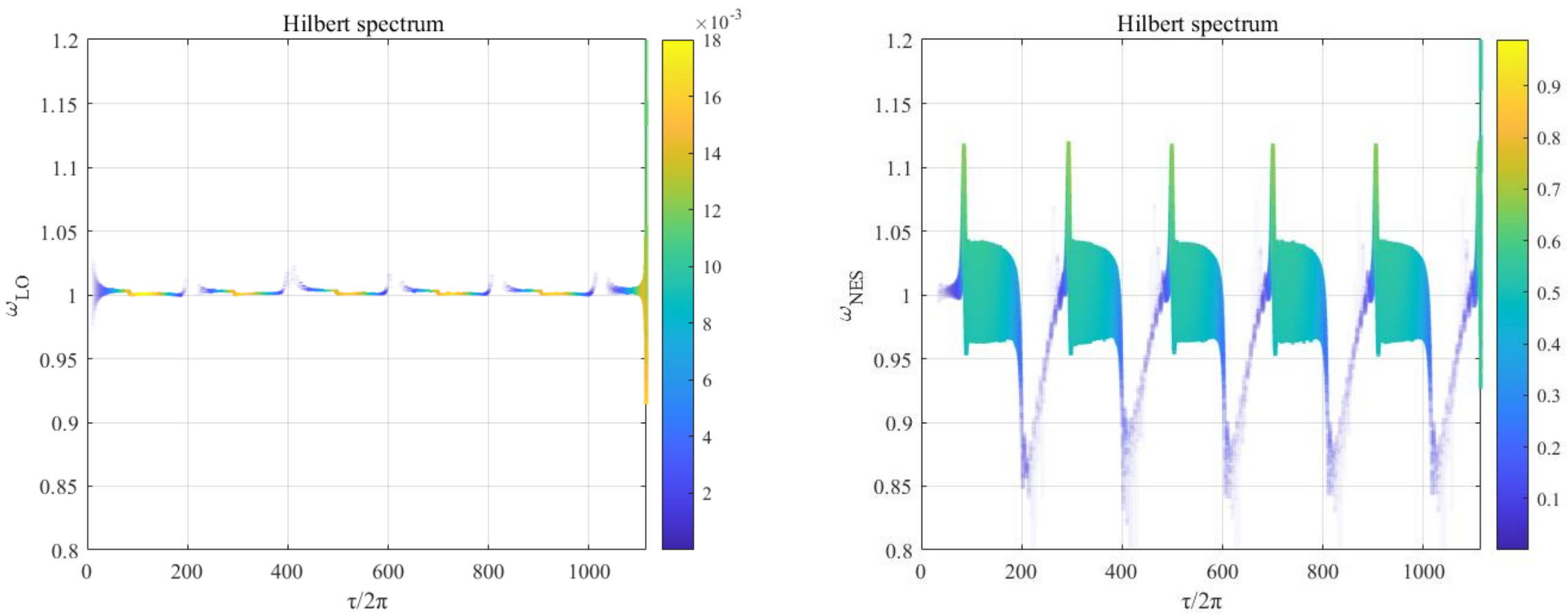

(a) (b)

Fig.11. Hilbert-Huang transform with the system parameters fixed at $\varepsilon$=0.001, $k$=1, $\lambda$=0.5, $k_2$=0.7, $\sigma$=4, $A$=0.5; (a) Hibert spectrum for the displacements $x_1$ of the LO; (b) Hibert spectrum for the displacements $x_2$ of the GNES.

*5.2. Evolution of oscillations on two-time-scale*

*5.2.1 Case 1*

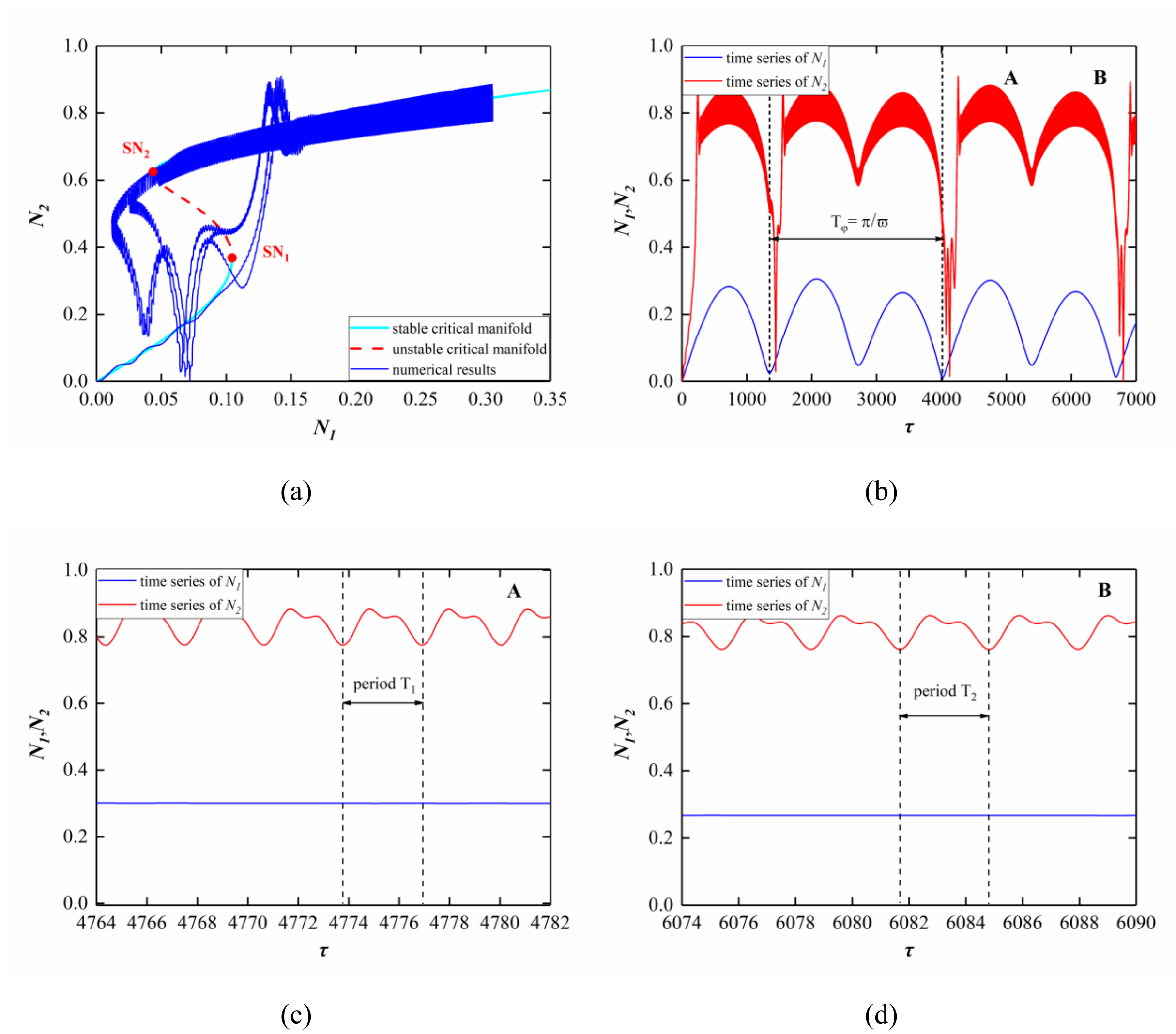

Fig.12. Numerical simulations with the system parameters fixed at $\varepsilon$=0.001, $k$=1, $\lambda$=0.5, $k_2$=0.7, $\sigma$=4, $A$=1.2;(a) Superposition of the critical manifold on the modulus plane ($N_1$, $N_2$) and corresponding numerical results;(b) Superposition of numerical results of the evolution of $N_1$ and $N_2$;(c) Close-up A; (d) Close-up B.

Fixing system parameters at $\varepsilon$=0.001, $k$=1, $\lambda$=0.5, $k_2$=0.7, $\sigma$=4, and $A$=1.2, still, the trajectory of the system starting from the origin moves along the lower branch and the upper branch of the critical manifold until it reaches $SN_1$ and $SN_2$ bifurcation points, in which the trajectory makes a jump from one stable critical manifold to the other stable one. The two jumping processes and the trajectories on two stable critical manifolds form a closed path. Under the action of external periodic excitation, this closed path exhibits periodic behavior. It is important to note, however, that except for the first closed path, the period of subsequently closed paths is twice that of the first closed path, since the trajectory moves twice along the upper branch of the stable critical manifold in each subsequently closed path. The detailed information is clear from subfigures (a)-(b) of Fig.12. Let us make two close-ups of the time series of the evolution of $N_1$ and $N_2$.

From the close-up A, shown in subfigure (c) of Fig.12, one can identify that $N_1$ remains unchanged during one oscillation period $T_1$ of $N_2$. The same situation occurring in the close-up B is described in subfigure (d) of Fig.12.

It is noteworthy that the periods of two movements of the trajectory on the upper branch of the stable critical manifold, in each subsequently closed path, are equal. That is $T_1=T_2 << T_\varphi$. Thereby, the oscillations in this situation are on a two-time scale. In the subfigure (a) of Fig.13, the Hilbert spectrum shows that $\omega_{LO}$ is nearly equal to 1 throughout the entire time range. In the subfigure (b) of Fig.13, we find that $\omega_{NES}$ oscillates rapidly, representing intense energy transfer, throughout the interval (0.95,1.05) in the entire time. Thus, one realizes intense modal interactions occurring in the system.

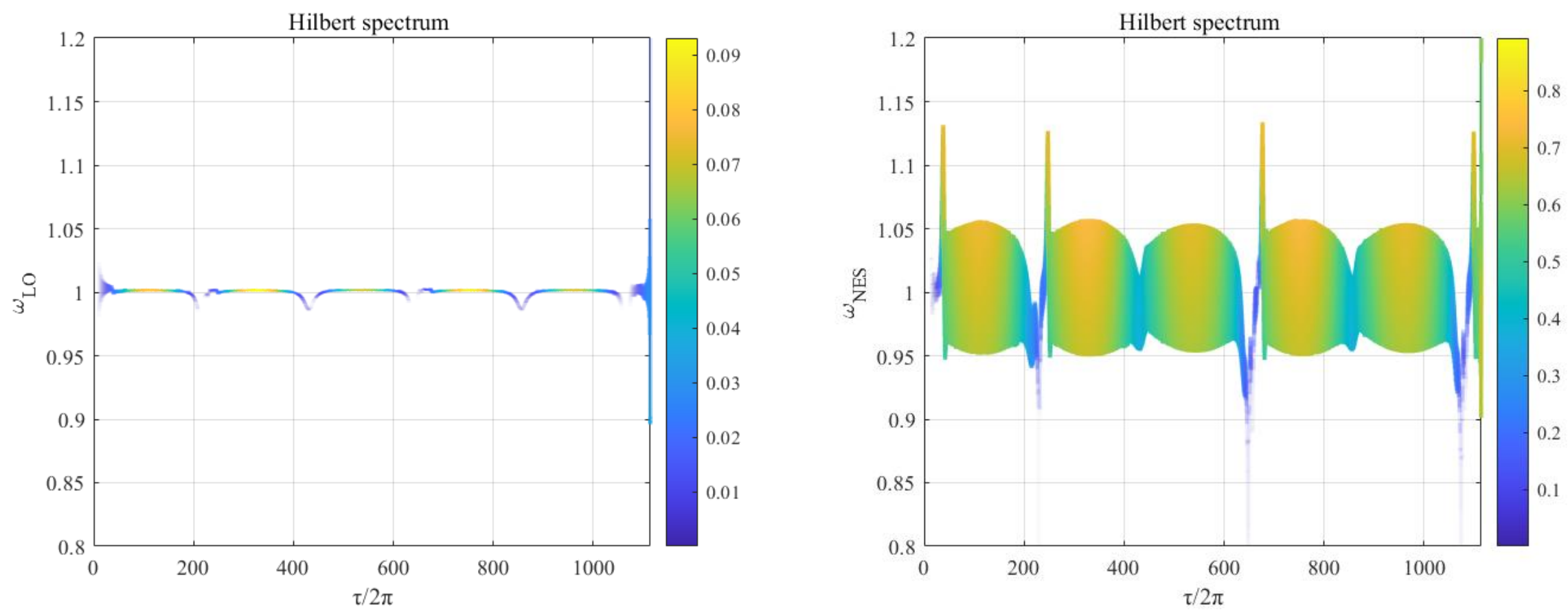

(a) (b)

Fig.13. Hilbert-Huang transform with the system parameters fixed at $\varepsilon$=0.001, $k$=1, $\lambda$=0.5, $k_2$=0.7, $\sigma$=4, $A$=1.2; (a) Hibert spectrum for the displacements $x_1$ of the LO; (b) Hibert spectrum for the displacements $x_2$ of the GNES.

### *5.2.2. Case 2*

We take the same values of the system parameters as in case 1. The trajectory on the critical manifold will make an obvious change when the amplitude of the external forcing increases to $A$=1.5. In fact, with the influence of the external forcing, the trajectory, starting from the origin, moves along the lower branch of the stable critical manifold until the $SN_1$ bifurcation point, and then, the trajectory jumps to the upper branch of the stable critical manifold and continues to perform periodic movements along it. It is shown in subfigures (a) and (b) of Fig.14.Viewing from subfigure (c) of Fig.14, one can find that $N_1$ remains unchanged during one oscillation period T of $N_2$. Due to $T << T_\varphi$, hence, we define that the oscillations in case (b) belong to the type of oscillations on a two-time-scale. Similarly, we use the two Hilbert spectrums corresponding to displacements for the system to display the time-frequency-energy relation diagrams. The diagrams reveal the energy transfer between different time scales, viewed in subfigures (a) and (b) of Fig.15.

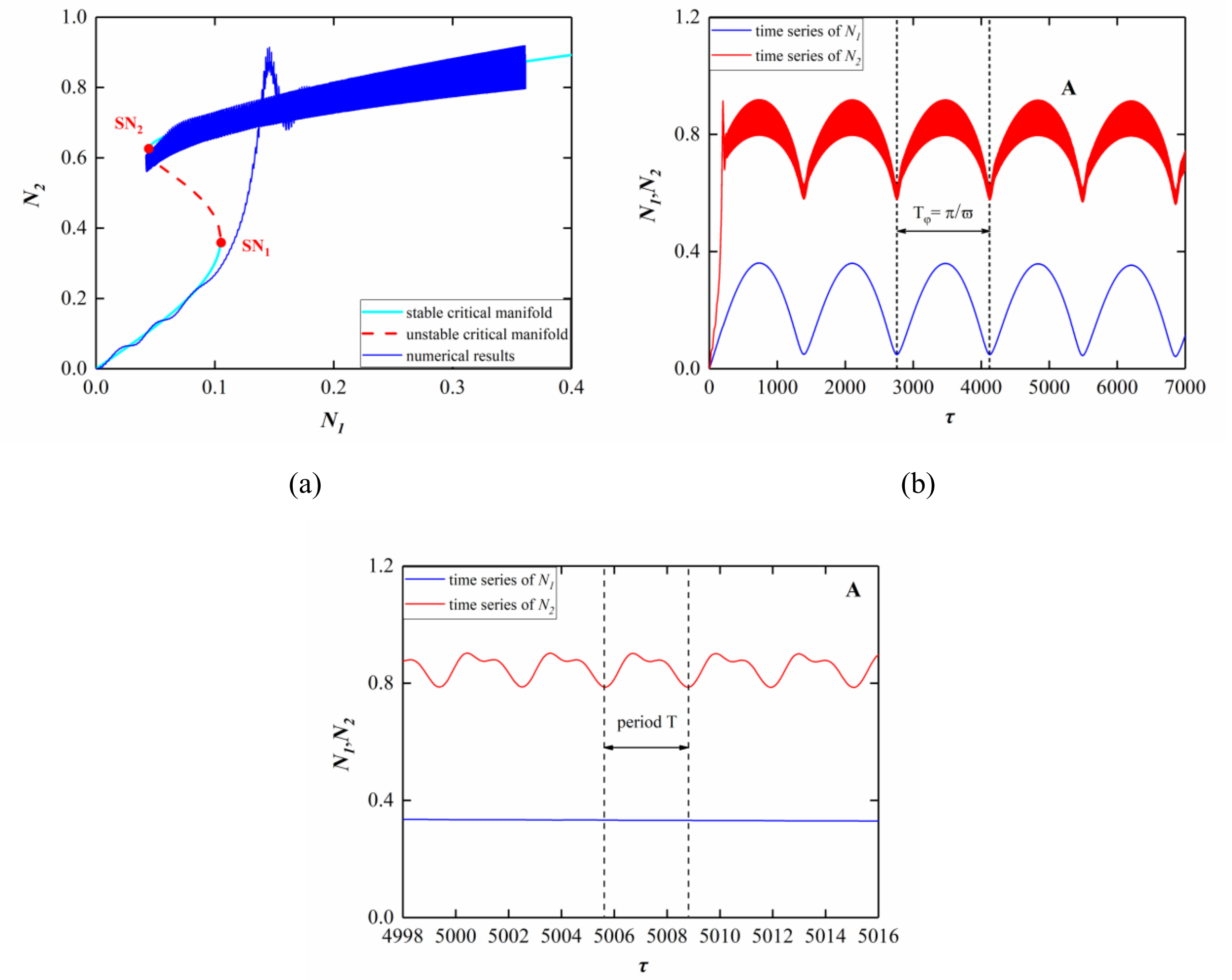

(a) (b)

(c)

Fig.14. Numerical simulations with the system parameters fixed at $\varepsilon$=0.001, $k$=1, $\lambda$=0.5, $k_2$=0.7, $\sigma$=4, $A$=1.5;(a) Superposition of the critical manifold on the modulus plane ($N_1$, $N_2$) and corresponding numerical results;(b) Superposition of numerical results of the evolution of $N_1$ and $N_2$;(c) Close-up A.

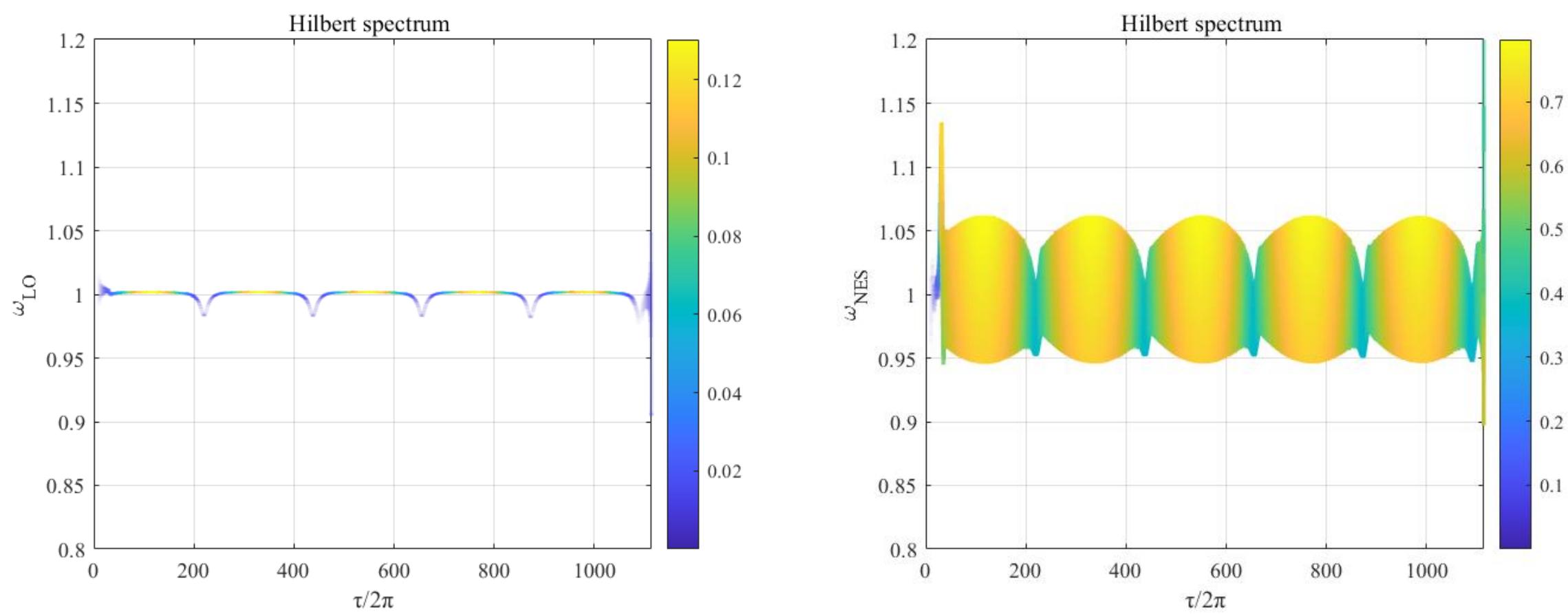

(a) (b)

Fig.15. Hilbert-Huang transform with the system parameters fixed at $\varepsilon$=0.001, $k$=1, $\lambda$=0.5, $k_2$=0.7, $\sigma$=4, $A$=1.5; (a) Hibert spectrum for the displacements $x_1$ of the LO; (b) Hibert spectrum for the displacements $x_2$ of the GNES.

*5.2.3. Case 3*

For case 3, the system parameters and the amplitude of external forcing are taken the values at $\varepsilon$=0.001, $k$=0.5, $\lambda$=0.5, $k_2$=0.5, $\sigma$=2, $A$=0.5. Starting from the origin the trajectory of the system moves along the lower branch of the critical manifold until it reaches the bifurcation point $SN_1$. Then, it jumps to the upper branch of the stable critical manifold and makes very fast oscillations. After that, the trajectory meets bifurcation point $SN_2$ and jumps to the lower branch of the stable critical manifold, in which the temporary damped oscillations occur. Subsequently, the trajectory moves along the lower branch until completing a closed path of one period.

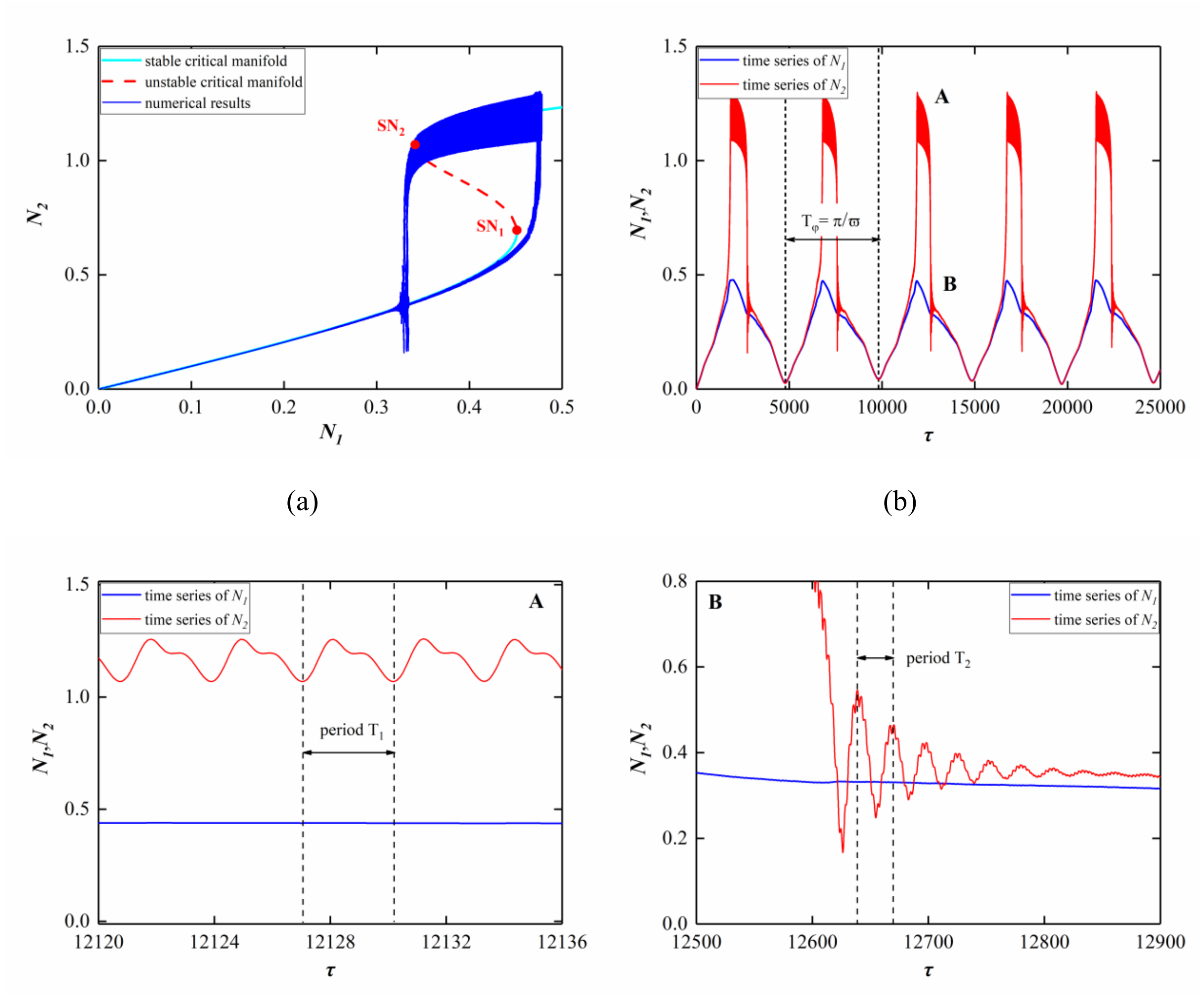

(c) (d)

Fig.16. Numerical simulations with the system parameters fixed at $\varepsilon$=0.001, $k$=0.5, $\lambda$=0.5, $k_2$=0.5, $\sigma$=2, $A$=0.5;(a) Superposition of the critical manifold on the modulus plane ($N_1$, $N_2$) and corresponding numerical results;(b) Superposition of numerical results of the evolution of $N_1$ and $N_2$;(c) Close-up A; (d) Close-up B.

When we make two close-ups of the time series of the evolution of $N_1$ and $N_2$, one can find that $N_1$ remains unchanged during one oscillation period $T_1$ of $N_2$ in subfigure (c) of Fig.16. However, in subfigure (d) of Fig.16, $N_2$ makes several temporary damped oscillations, and then it moves almost along the path of $N_1$. Due to the very short occurrence time of the temporary damped oscillations, we recognize that there are only oscillations on a two-time scale in this case, where $T_1 << T_\varphi$. We give the corresponding Hilbert spectrums to exhibit the time-frequency-energy relation between the displacements for the LO and the GNES. It is shown in Fig.17.

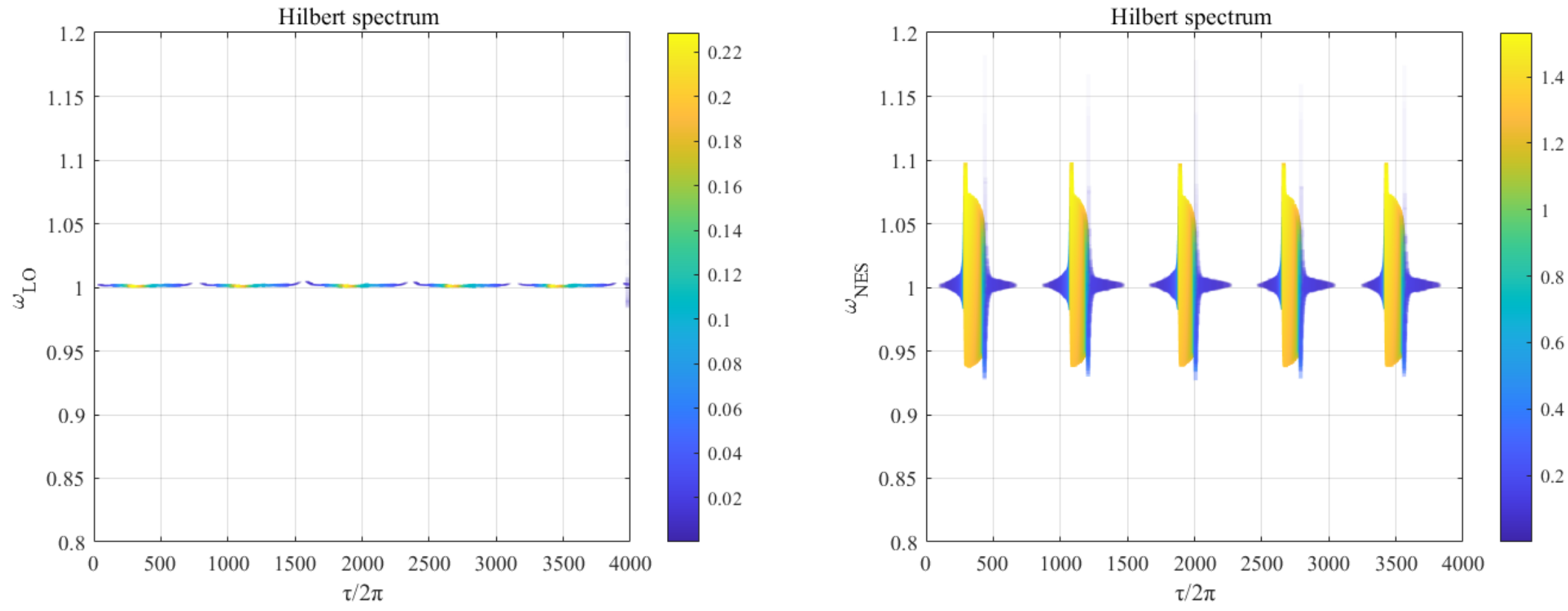

(a) (b)

Fig.17. Hilbert-Huang transform with the system parameters fixed at $\varepsilon$=0.001, $k$=0.5, $\lambda$=0.5, $k_2$=0.5, $\sigma$=2, $A$=0.5; (a) Hibert spectrum for the displacements $x_1$ of the LO; (b) Hibert spectrum for the displacements $x_2$ of the GNES.

### *5.3. Evolution of oscillations on single-time-scale*

For the parameters fixed at $\varepsilon$=0.01, $k$=1, $\lambda$=1.5, $k_2$=0.2, $\sigma$=1, and $A$=0.27, one can notice the lower branch of the stable critical manifold is represented as a diagonal line in the modulus plane ($N_1$, $N_2$). Still, make the origin the initial point. Indeed, the trajectory moves back and forth along the critical manifold in a periodic way under the excitation of external forcing, as seen in subfigure (a) of Fig.18. Viewing from the superposition of numerical results of the evolution of $N_1$ and $N_2$, they oscillate periodically with the same period, i.e., $T=T_\varphi$. This result is depicted in subfigure (b) of Fig.18. Consequently, we regard the oscillations as oscillations on a single time scale.

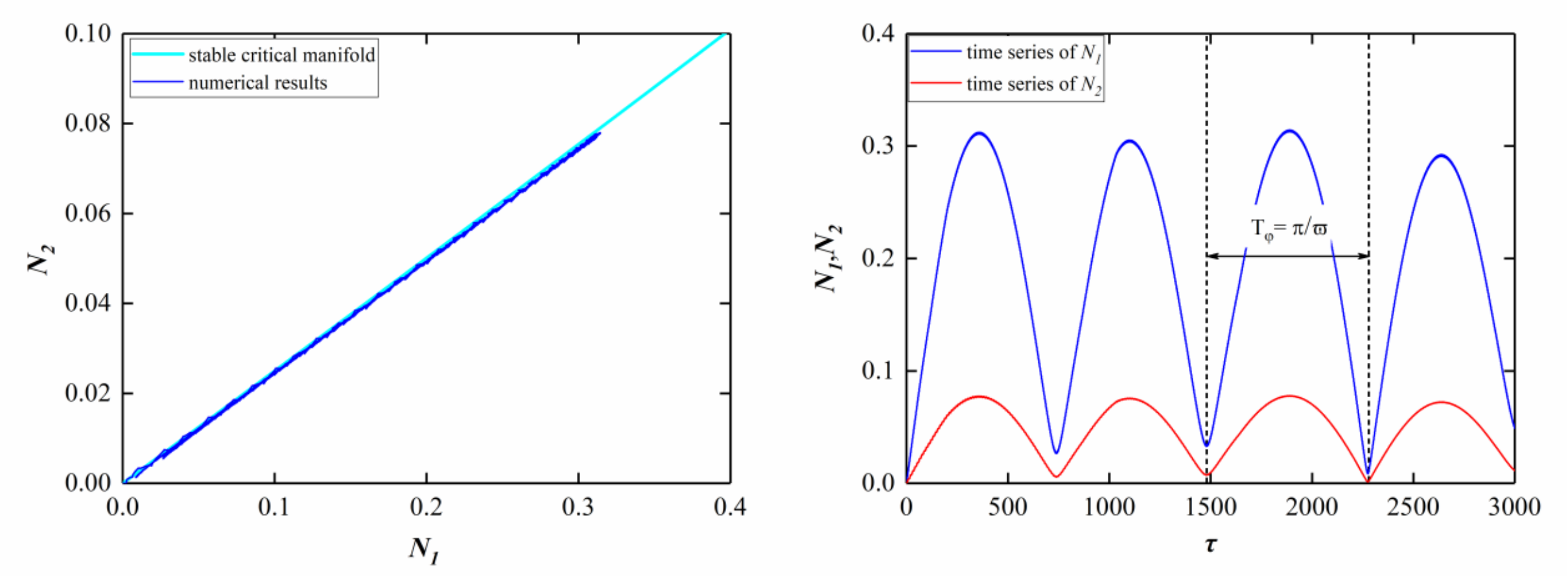

(a) (b)

Fig.18. Numerical simulations with the system parameters fixed at $\varepsilon$=0.01, $k$=1, $\lambda$=1.5, $k_2$=0.2, $\sigma$=1, $A$=0.27;(a)

Superposition of the critical manifold on the modulus plane ($N_1$, $N_2$) and corresponding numerical results;(b) Superposition of numerical results of the evolution of $N_1$ and $N_2$.

In subfigures (a) and (b) of Fig.19, the Hilbert spectrums corresponding to displacements $x_1$ and $x_2$ show that there may be weak energy transfer between the LO and the GNES. We can find that the instantaneous frequency $\omega_{\mathrm{LO}}$ and $\omega_{\mathrm{NES}}$ are both located near 1, and their colors are slowly changing.

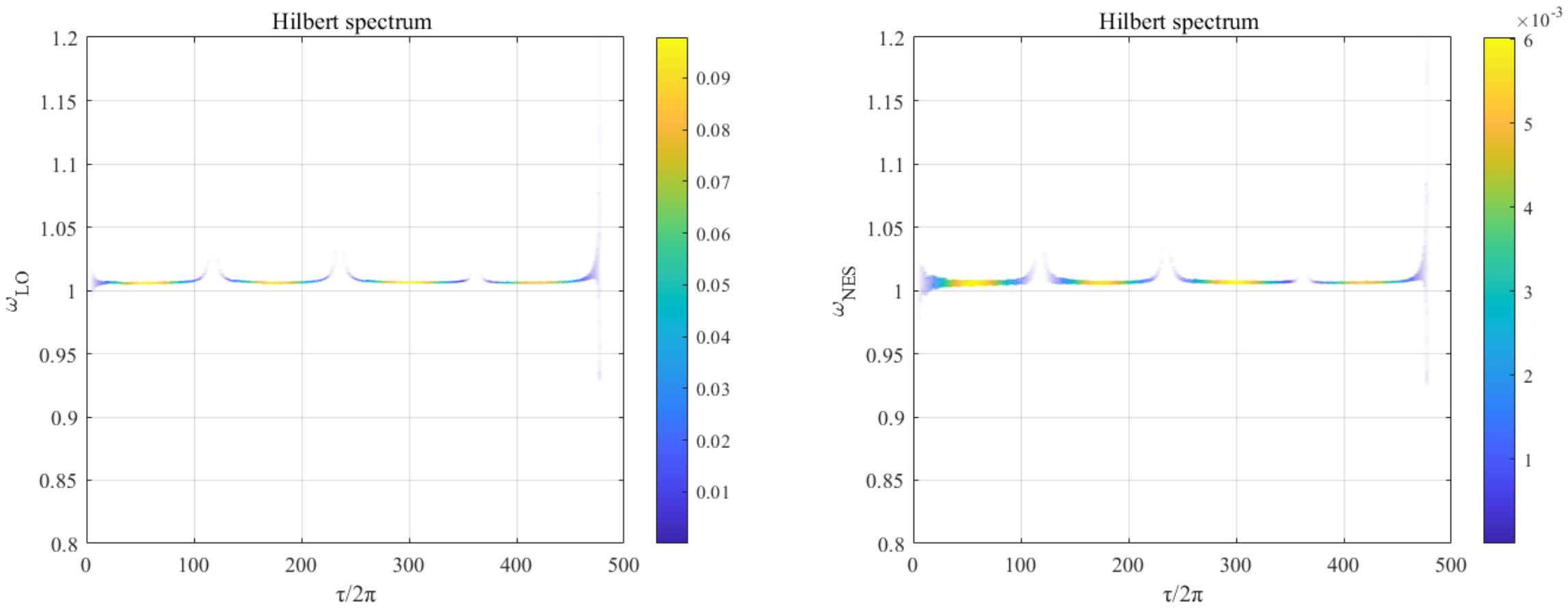

(a) (b)

Fig.19. Hilbert-Huang transforms with the system parameters fixed at $\varepsilon$=0.01, $k$=1, $\lambda$=1.5, $k_2$=0.2, $\sigma$=1, $A$=0.27; (a) Hibert spectrum for the displacements $x_1$ of the LO; (b) Hibert spectrum for the displacements $x_2$ of the GNES.

### *5.4. Oscillations in other situations*

In this subsection, we will discuss two special types of oscillations, which are completely different from the types discussed above.

#### *5.4.1. Point-type oscillations*

In subfigure (a) of Fig.20, we notice that the critical manifold predicts an approximation of failure, as the trajectory no longer moves along the critical manifold. The trajectory exhibits a spiral motion and ultimately tends towards the vicinity of a point. Because motion has not been found to contain significant differences on multiple time scales, we here define the motion as "point-type oscillations". It is worth noting that this type of oscillation indicates almost no change in the amplitude of the LO and the GNES, implying that it is a stationary oscillation of the original system (2). The corresponding Hilbert spectrums, depicted in subplots (a) and (b) of Fig.21, verify that there may be no type of energy transfer in the system except for the initial effluxion of time since the instantaneous energies of displacements for the LO and the GNES remain unchanged. In Fig.21, we notice that the instantaneous frequency $\omega_{\mathrm{LO}}$ of the LO is close to 1, while the instantaneous frequency $\omega_{\mathrm{NES}}$ of the GNES oscillates in the vicinity of 1.

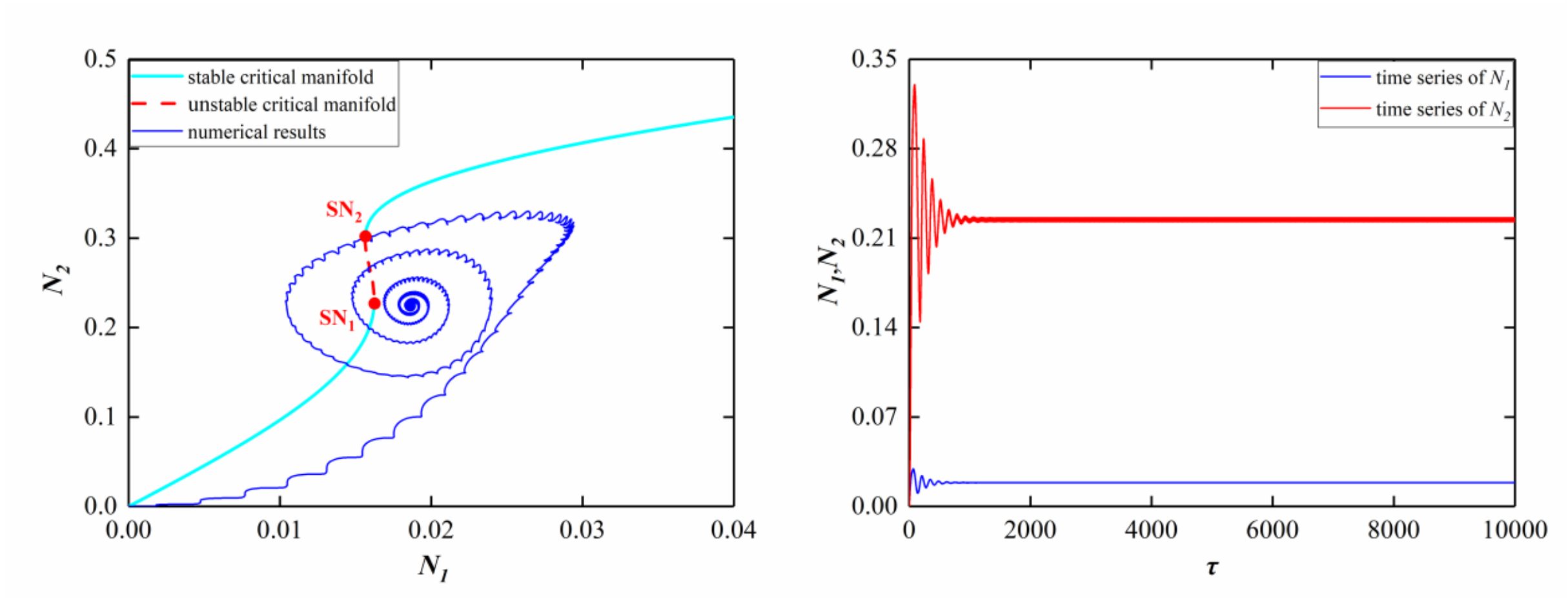

(a) (b)

Fig.20. Numerical simulations with the system parameters fixed at $\varepsilon$=0.01, $k$=1, $\lambda$=1.5, $k_2$=0.9, $\sigma$=0.5, $A$=0.2;(a) Superposition of the critical manifold on the modulus plane ($N_1$, $N_2$) and corresponding numerical results;(b) Superposition of numerical results of the evolution of $N_1$ and $N_2$.

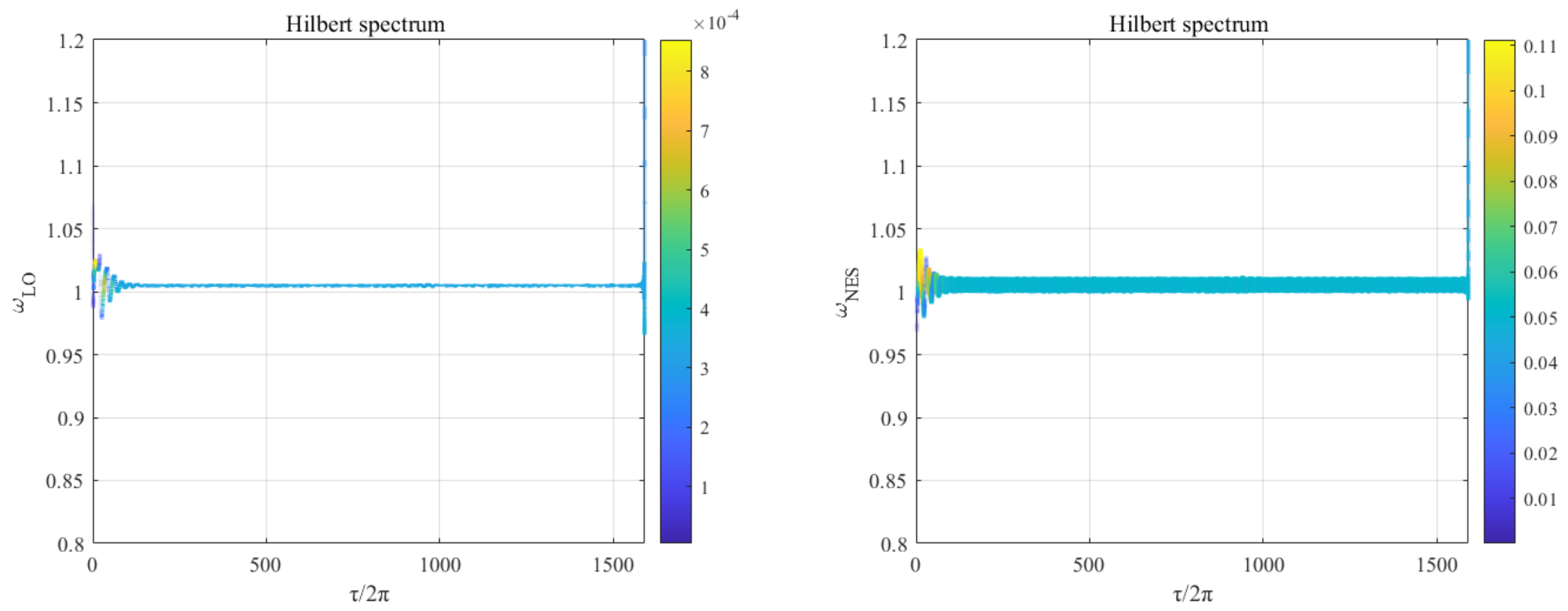

(a) (b)

Fig.21. Hilbert-Huang transforms with the system parameters fixed at $\varepsilon$=0.01, $k$=1, $\lambda$=1.5, $k_2$=0.9, $\sigma$=0.5, $A$=0.2; (a) Hibert spectrum for the displacements $x_1$ of the LO; (b) Hibert spectrum for the displacements $x_2$ of the GNES.

### *5.4.2. Ring-type oscillations*

Similarly, the critical manifold forecasts an approximation of failure in this case, as seen in subfigure (a) of Fig.22. The trajectory ultimately tends toward the vicinity of a ring. Viewing from subfigures (b) and (c) of Fig.22, the numerical results of the evolution of $N_1$ and $N_2$ can be realized, in which both of them demonstrate periodic oscillation by the same period, i.e., $T_1$=$T_2$. However, on the time series of the amplitude of $N_2$, there are many tiny oscillations, depicted in subfigure (d) of Fig.22. For the sake of intuition, we define the motion of this case as “ring-type oscillations” because the tiny oscillations on the amplitude of $N_2$ are so very small that we can neglect its influence. In subfigures (a) and (b) of Fig.23, we notice that the instantaneous frequencies $\omega_{LO}$ and $\omega_{NES}$ oscillate at nearly the same frequency interval at the same rate, suggesting the occurrence of energy exchange between the LO and the GNES.

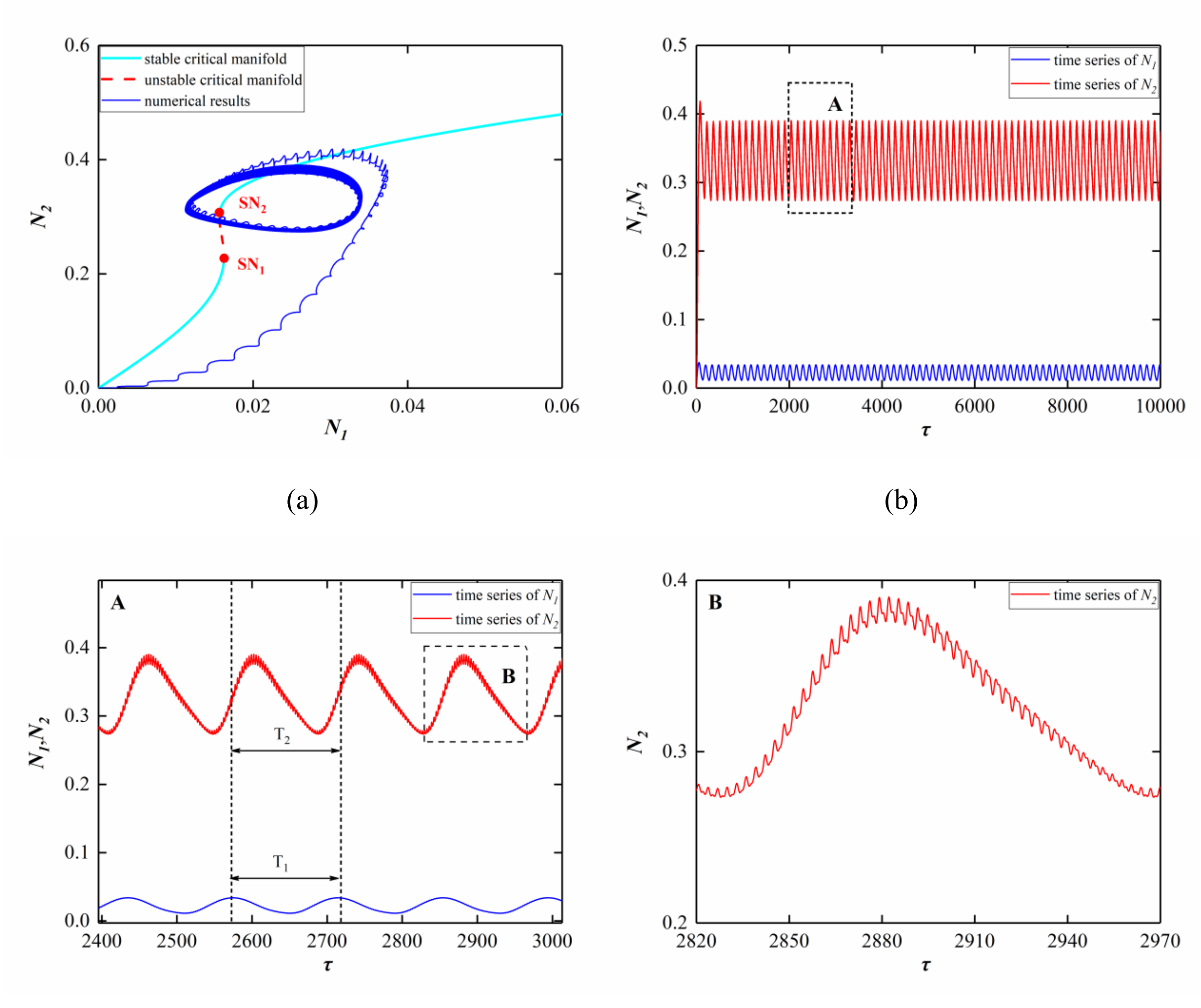

(a) (b)

(c) (d)

Fig.22. Numerical simulations with the system parameters fixed at $\varepsilon$=0.01, $k$=1, $\lambda$=1.5, $k_2$=0.9, $\sigma$=1, $A$=0.27;(a) Superposition of the critical manifold on the modulus plane ($N_1$, $N_2$) and corresponding numerical results;(b) Superposition of numerical results of the evolution of $N_1$ and $N_2$;(c) Close-up A; (d) Close-up B.

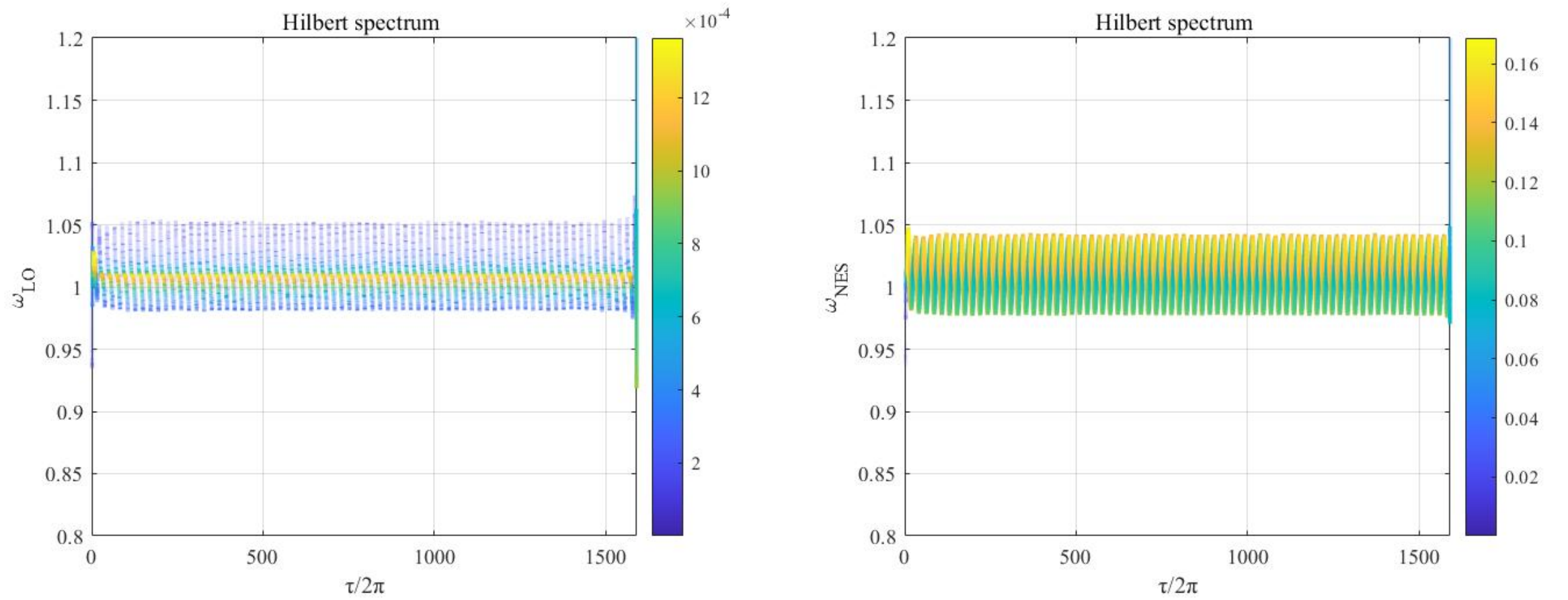

(a) (b)

Fig.23. Hilbert-Huang transforms with the system parameters fixed at $\varepsilon$=0.01, $k$=1, $\lambda$=1.5, $k_2$=0.9, $\sigma$=1, $A$=0.27; (a) Hibert spectrum for the displacements $x_1$ of the LO; (b) Hibert spectrum for the displacements $x_2$ of the GNES.

## 6. Conclusions and discussions

The results presented in the above sections demonstrate that the two-DOF system, consisting of a LO and a GNES, occurs modal interactions on multiple time scales. Indeed, around the neighbor of the 1:1:1 resonance regime, we find the fast-slow forms of the slow flow, in which the derivative of complex amplitude $\varphi_1$ with respect to time is far less than that of complex amplitude $\varphi_2$ to time. Therefore, the complex amplitude $\varphi_1$ is realized as a generalized parameter because of its slowly-varying property. Additionally, due to the fact $\varepsilon$ is a very small parameter, the fast subsystem and slow subsystem are considered as the undisturbed system to analyze the dynamics of the system approximatively.

According to that, we acquire the critical manifold, projected on the modulus square plane, to capture the dynamical behavior of the system. With the change of $Z_1$, we find that the critical manifold on the modulus square plane contains folding structures, consisting of a pair of SN bifurcation points, the stable critical manifolds, and the unstable one, which satisfies the condition $\alpha_2^2$-3$\alpha_1\alpha_3$>0. When system parameters meet the condition $\alpha_2^2$-3$\alpha_1\alpha_3$=0, the unstable critical manifold disappears. At the same time, the TP point occurs at the tangent of the upper branch and the lower branch of the stable critical manifold. Furthermore, as $\alpha_2^2$-3$\alpha_1\alpha_3$<0, the lower branch vanishes, demonstrating that there is only a single stable fixed point. All these distinct structures induce rich patterns of the modal interactions between the LO and the GNES. By using the expression of polar coordinates, we get detailed position information of fixed points in the above three cases.

The results of numerical simulations show the oscillations on multiple time scales, resulting from the modal interactions between the LO and the GNES. Indeed, as the system parameters and amplitude of external excitation are fixed at different values, we find that oscillations on the three-time scale, oscillations on the two-time scale, and oscillations on the single-time scale can be captured by the critical manifold approximatively. Furthermore, we define and discuss, in detail, distinct patterns of these oscillations. In addition, two additional types of oscillations, namely point-type oscillations and ring-type oscillations, are detected. It is worth noting that the two types of oscillations cannot be predicted from the critical manifold. Finally, by using the HHT, we get the time-frequency-energy relation of these oscillations and discuss the energy transfers on different time scales.

The multiscale energy transfers probably imply the diverse vibration absorption efficiencies. It can be intentionally realized for enhancing efficiencies of GNES devices for vibration passive control. However, we here would like to point out that the discussion on the design parameters of GNES in this article is limited to theoretical analysis, and its application requires further analysis and exploration. Therefore, the design of GNES suitable for different vibration scenarios will be considered in future work.

## CRediT authorship contribution statement

**Lan Huang:** Conceptualization, Methodology, Investigation, Writing – original draft.

**Shufeng Lu:** Conceptualization, Methodology, Validation, Writing – review & editing.

**Wensai Ma:** Conceptualization, Software, Validation, Writing – review & editing.

## Declaration of Competing Interest

The authors declare that they have no known competing financial interests or personal relationships that could have appeared to influence the work reported in this paper.

## Data Availability

Data will be made available on reasonable request.

## Acknowledgment

This work was supported by the National Natural Science Foundation of China (No.12172182), Natural Science Foundation of Inner Mongolia (2023JQ14), Inner Mongolia Autonomous Region Science and Technology Plan Project (2022JBGS0041), and Research Foundation of Inner Mongolia University of Technology (DC2500000692).

There is a preprint [47] of this work available for review.

## References

[1] Weaver Jr W, Timoshenko SP, Young DH. Vibration problems in engineering: John Wiley & Sons; 1991.

[2] Kim Y, Park J. A theory for the free vibration of a laminated composite rectangular plate with holes in aerospace applications. Composite Structures. 2020;251:112571.

[3] Lu S, Hu R, Qiao H, Yang Y, Zhang W, Song X. Dynamic analysis of carbon fiber-wound reinforced composite truncated conical shell under thermal conditions: Theory and experimental validation. Thin-Walled Structures. 2024;205:112496.

[4] Wang J, Chang Z, Cao G, Lai S-K. Predicting delamination in composite laminates through semi-analytical dynamic analysis and vibration-based quantitative assessment. Thin-Walled Structures. 2024;204:112346.

[5] Ding H, Chen L-Q. Designs, analysis, and applications of nonlinear energy sinks. Nonlinear Dynamics. 2020;100:3061-107.

[6] Li Z, Wang Y, Qiao B, Wen B, Chen X. Experimental investigation of aeroelastic instabilities in an aeroengine fan: Using acoustic measurements. Aerospace Science and Technology. 2022;130:107927.

[7] Yang Y, Chen G, Ouyang H, Yang Y, Cao D. Nonlinear vibration mitigation of a rotor-casing system subjected to imbalance–looseness–rub coupled fault. International Journal of Non-Linear Mechanics. 2020;122:103467.

[8] Wu Z, Hao L, Zhao W, Ma Y, Bai S, Zhao Q. Modeling and vibration analysis of an aero-engine dual-rotor-support-casing system with inter-shaft rub-impact. International Journal of Non-Linear Mechanics. 2024;165:104757.

[9] Huang J, Yang Y, Peng B, Li S. Vibration and Fault Analysis of a Rotor System of a Twin-Spool Turbo-Jet Engine in Ground Test. Aerospace. 2024;11:724.

[10] Xu Z, Wei J, Wei H, Liu Z, Zhang Y. A novel resonance attractor evaluation method for traveling wave vibration of thin-walled gears: Modeling and experiments. Journal of Sound and Vibration. 2024;569:118081.

[11] Guo W, Xu Y, Liu Q, Lenci S, Li G. Reliability of hypersonic airfoil with freeplay and stochasticity via nonlinear energy sink. AIAA Journal. 2024;62:3258-70.

[12] Onoda J, Sano T, Kamiyama K. Active, passive, and semiactive vibration suppression by stiffness variation. AIAA Journal. 1992;30:2922-9.

[13] Xing Z-Y, Yang X-D. A combined vibration isolation system with quasi-zero stiffness and dynamic vibration absorber. International Journal of Mechanical Sciences. 2023;256:108508.

[14] Geng X, Ding H, Jing X, Mao X, Wei K, Chen L. Dynamic design of a magnetic-enhanced nonlinear energy sink. Mechanical Systems and Signal Processing. 2023;185:109813.

[15] Tang Y, Gao C, Li M, Ding Q. Novel active-passive hybrid piezoelectric network for vibration suppression in fluid-conveying pipes. Applied Mathematical Modelling. 2023;117:378-98.

[16] Ormondroyd J, Den Hartog J. The theory of the dynamic vibration absorber. Journal of Fluids Engineering. 1928;49.

[17] Gendelman O, Manevitch LI, Vakakis AF, M'Closkey R. Energy pumping in nonlinear mechanical oscillators: Part I - Dynamics of the underlying Hamiltonian systems. Journal of Applied Mechanics-Transactions of the Asme. 2001;68:34-41.

[18] Vakakis AF, Gendelman O. Energy pumping in nonlinear mechanical oscillators: Part II - Resonance capture. Journal of Applied Mechanics-Transactions of the Asme. 2001;68:42-8.

[19] Wang T, Tang Y, Yang T, Ma Z-S, Ding Q. Bistable enhanced passive absorber based on integration of nonlinear energy sink with acoustic black hole beam. Journal of Sound and Vibration. 2023;544:117409.

[20] Wang T, Tang Y, Qian X, Ding Q, Yang T. Enhanced nonlinear performance of nonlinear energy sink under large harmonic excitation using acoustic black hole effect. Nonlinear Dynamics. 2023;111:12871-98.

[21] Vakakis AF, Gendelman OV, Bergman LA, McFarland DM, Kerschen G, Lee YS. Nonlinear targeted energy transfer in mechanical and structural systems: Springer Science & Business Media; 2008.

[22] Arnold V, Kozlov V, Neishtadt A. Dynamical Systems III, Encyclopaedia of Mathematical Sciences, Vol. 3. Springer, Berlin; 1988.

[23] McFarland DM, Bergman LA, Vakakis AF. Experimental study of non-linear energy pumping occurring at a single fast frequency. International Journal of Non-Linear Mechanics. 2005;40:891-9.

[24] Kerschen G, Gendelman O, Vakakis AF, Bergman LA, McFarland DM. Impulsive periodic and quasi-periodic orbits of coupled oscillators with essential stiffness nonlinearity. Communications in Nonlinear Science and Numerical Simulation. 2008;13:959-78.
[25] Lee YS, Vakakis AF, Bergman LA, McFarland DM, Kerschen G. Suppression aeroelastic instability using broadband passive targeted energy transfers, part 1: Theory. AIAA Journal. 2007;45:693-711.
[26] Lee YS, Kerschen G, McFarland DM, Hill WJ, Nichkawde C, Strganac TW, et al. Suppressing aeroelastic instability using broadband passive targeted energy transfers, part 2: experiments. AIAA Journal. 2007;45:2391-400.
[27] Lee YS, Vakakis AF, Bergman LA, McFarland DM, Kerschen G. Enhancing the robustness of aeroelastic instability suppression using multi-degree-of-freedom nonlinear energy sinks. AIAA Journal. 2008;46:1371-94.
[28] Tian W, Li Y, Li P, Yang Z, Zhao T. Passive control of nonlinear aeroelasticity in hypersonic 3-D wing with a nonlinear energy sink. Journal of Sound and Vibration. 2019;462:114942.
[29] Wall I, Amoozgar M, Popov A. Aeroelasticity of an aircraft wing with nonlinear energy sink. Aerospace Science and Technology. 2024;155:109684.
[30] Huang C, Zheng G, Nie X, Yang G. Supercritical and subcritical aeroelastic behaviors of a three-dimensional wing coupled with a nonlinear energy sink. International Journal of Non-Linear Mechanics. 2024;161:104692.
[31] Bergeot B, Bellizzi S, Cochelin B. Analysis of steady-state response regimes of a helicopter ground resonance model including a non-linear energy sink attachment. International Journal of Non-Linear Mechanics. 2016;78:72-89.
[32] Bergeot B, Bellizzi S, Cochelin B. Passive suppression of helicopter ground resonance using nonlinear energy sinks attached on the helicopter blades. Journal of Sound and Vibration. 2017;392:41-55.
[33] Miao X, He J, Jiang D, Zhang D, Pennacchi P, Fei Q. Nonlinear vibration suppression of a dual-rotor system with combined misalignment using the nonlinear energy sink. Chaos, Solitons & Fractals. 2025;192:116048.
[34] Yao H, Cao Y, Ding Z, Wen B. Using grounded nonlinear energy sinks to suppress lateral vibration in rotor systems. Mechanical Systems and Signal Processing. 2019;124:237-53.
[35] Geng X-F, Ding H, Mao X-Y, Chen L-Q. A ground-limited nonlinear energy sink. Acta Mechanica Sinica. 2022;38:521558.
[36] Sui P, Shen Y, Wang X. Study on response mechanism of nonlinear energy sink with inerter and grounded stiffness. Nonlinear Dynamics. 2023;111:7157-79.
[37] Li S, Zhou X, Chen J. Hamiltonian dynamics and targeted energy transfer of a grounded bistable nonlinear energy sink. Communications in Nonlinear Science and Numerical Simulation. 2023;117:106898.
[38] Weiss M, Savadkoohi AT, Gendelman OV, Lamarque C-H. Dynamical behavior of a mechanical system including Saint-Venant component coupled to a non-linear energy sink. International Journal of Non-Linear Mechanics. 2014;63:10-8.
[39] Dekemele K, Habib G. Inverted resonance capture cascade: modal interactions of a nonlinear energy sink with softening stiffness. Nonlinear Dynamics. 2023;111:9839-61.
[40] Charlemagne S, Lamarque C-H, Savadkoohi AT. Dynamics and energy exchanges between a linear oscillator and a nonlinear absorber with local and global potentials. Journal of Sound and Vibration. 2016;376:33-47.
[41] Bergeot B. Effect of stochastic forcing on the dynamic behavior of a self-sustained oscillator coupled to a non-linear energy sink. International Journal of Non-Linear Mechanics. 2023;150:104351.
[42] Bergeot B, Berger S. Fast–slow analysis of passive mitigation of self-sustained oscillations by means of a bistable nonlinear energy sink. Physica D: Nonlinear Phenomena. 2024;460:134063.
[43] Huang L, Yang X. Multiscale energy transfers in a forced two-DOF oscillatory system with a grounded nonlinear energy sink. arXiv preprint arXiv:241109931. 2024.
[44] Manevitch LI. The description of localized normal modes in a chain of nonlinear coupled oscillators using complex variables. Nonlinear Dynamics. 2001;25:95-109.
[45] Fenichel N. Geometric singular perturbation theory for ordinary differential equations. Journal of Differential Equations. 1979;31:53-98.
[46] Huang NE, Wu Z. A review on Hilbert-Huang transform: Method and its applications to geophysical studies. Reviews of Geophysics. 2008;46.
[47] Huang L, Lu S, Ma W. Modal interactions between a linear oscillator and a nonlinear absorber: multiscale energy

transfers. arXiv preprint arXiv:2401.16869.